\documentclass[english,twocolumn,showpacs,prb,superscriptaddress]{revtex4}
\usepackage{epsfig}
\usepackage{amsmath}
\usepackage{graphicx}
\usepackage{babel}
\newcommand{\dg}{^{\dagger }}
\newcommand{\si}{\sigma}
\newcommand{\rarrow}{\rightarrow}
\newcommand{\bk}{{\bf k}}
\newcommand{\bq}{{\bf q}}
\newcommand{\bp}{{\bf p}}
\newcommand{\bQ}{{\bf {Q}}}
\newcommand{\bx}{{\bf{x}}}
\newcommand{\by}{{\bf{y}}}
\newcommand{\ba}{{\bf{a}}}
\newcommand{\br}{{\bf{r}}}
\newcommand{\bX}{{\bf{X}}}
\newlength{\figwidth}
\newcommand{\fg}[3]
{
\begin{figure}[ht]

\vspace*{-0cm}
\[
\includegraphics[width=\figwidth]{#1}
\]
\vskip -0.2cm
\caption{\label{#2}
\small#3
}
\end{figure}}

\begin{document}
\title{Itinerancy and Hidden Order in $URu_2Si_2$}
\author{V. Tripathi}
\affiliation
{Theory of Condensed Matter Group, Cavendish Laboratory,
University of Cambridge, Madingley Road, Cambridge CB3 0HE, United Kingdom}
\author{P. Chandra}
\affiliation
{Materials Theory Group,
Rutgers University, Piscataway, NJ 08855, U.S.A.}  
\author{P. Coleman}
\affiliation
{Materials Theory Group,
Rutgers University, Piscataway, NJ 08855, U.S.A.} 
\date{\today}
\begin{abstract}
We argue that key characteristics of the enigmatic transition at $T_0 = 17.5K $
in $URu_2Si_2$ indicate that the hidden order is a density wave formed
within a band of composite quasiparticles, whose detailed
structure is determined by local physics.
We expand on our proposal (with J.A. Mydosh) of the hidden order as
incommensurate orbital antiferromagnetism and present experimental
predictions to test our ideas.  We then turn towards a microscopic description of
orbital antiferromagnetism, exploring possible particle-hole pairings
within the context of a simple one-band model.  We end with
a discussion of recent high-field and thermal transport experiment,
and discuss their implications for the nature of the hidden order.
\end{abstract}

\maketitle

\section{Introduction}
The possibility of exotic particle-hole pairing leading to
quadrupolar and orbital charge
currents has been discussed extensively in the context of the two-dimensional 
Hubbard model.\cite{Halperin68,Affleck88,Kotliar88,Nersesyan89,Schulz89}
More recently d-wave charge-density wave states, both 
ordered\cite{Chakravarty01}
and fluctuating,\cite{Lee04} have been proposed
to explain the pseudogap phase in the underdoped 
cuprates and ground-states of doped two-leg Hubbard and $t-J$ 
ladders.\cite{Fjaerestad02,Wu04}
In this paper we discuss related anisotropic particle-hole pairing
in a different setting, namely that of three-dimensional Fermi
liquids. We believe that such pairing may occur in
the heavy fermion metal $URu_{2}Si_{2}$, and here we provide theoretical
support for our earlier publications (with J.A. Mydosh) on 
this topic.\cite{Chandra02a,Chandra02b,Mydosh03,Chandra03} 
Though the initial motivation for our orbital 
antiferromagnetism (OAFM) proposal
in $URu_2Si_2$ was primarily experimental, here we observe that 
coexistence of large electron-electron
repulsion and antiferromagnetic fluctuations favours node formation 
in particle-hole pairing and hence the formation of anisotropic
charge-density wave states.
After presenting technical details behind specific predictions
for neutron scattering and for NMR, 
we turn towards a microscopic description
of orbital antiferromagnetism.
We start by presenting the generalised
Landau parameters associated with this anisotropic pairing.
Next we study a toy model where this
instability occurs.  We end with a discussion of these results in
the light of more recent measurements, and also suggest further
experiments to test our ideas.

\figwidth=5cm
\fg{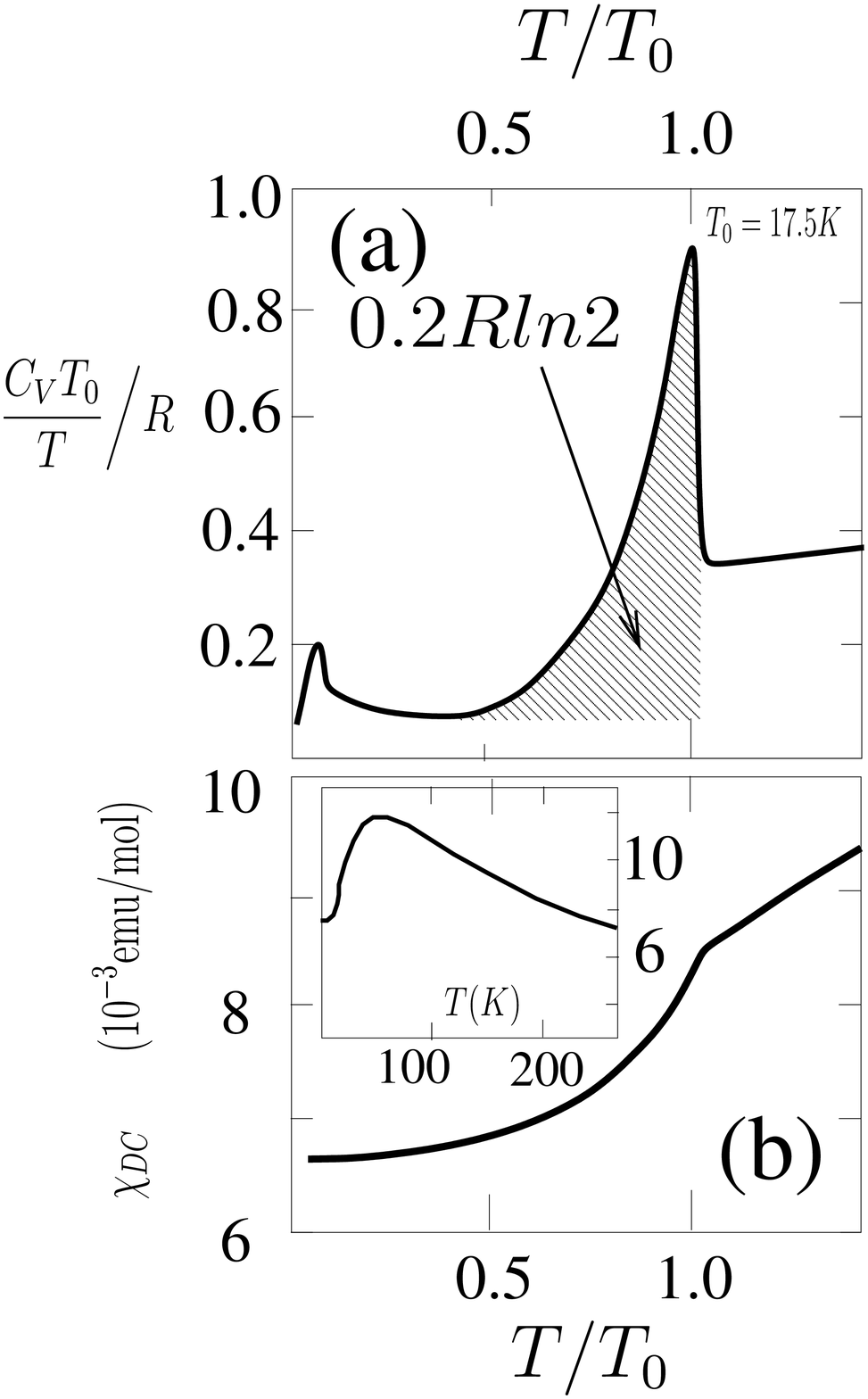}{fig1}
{\small 
Schematics of the (a) specific heat anomaly.  Data points taken from
Figs. 1.  of reference [13].  (b) The measured magnetic susceptibility
from reference [15]
Inset showing cross-over from high temperature local Curie behaviour
to low temperature Fermi liquid behaviour taken from Fig 2 of reference
[13]. 
}
\figwidth=7cm

The heavy fermion metal $URu_2Si_2$ displays a classic second-order
phase transition (see Figure 1) at
$T_0 = 17.5K$, and yet the nature of the associated order parameter
remains elusive nearly two decades after its discovery.  This 
phase transition is characterised by a large entropy loss\cite{Palstra85}
and sharp anomalies in the linear\cite{Palstra85} and the nonlinear 
susceptibilities,\cite{Miyako91,Ramirez92} the thermal 
expansion,\cite{deVisser86} and the resistivity,\cite{Palstra86} 
where standard mean-field relations between measured thermodynamic 
quantities are satisfied.\cite{Chandra94}
At the transition, neutron scattering experiments observe
gapped, propagating magnetic 
excitations\cite{Walter86,Mason91,Broholm87,Broholm91} that suggest
the formation of a spin density wave.
However, subsequent neutron scattering
measurements\cite{Broholm87,Broholm91} indicate that 
the staggered magnetic 
moment ($m_0 = 0.03\mu_B$ per U atom), 
is too small to account for the entropy loss at the 
transition,\cite{Buyers96} which has been attributed to
the development of an enigmatic hidden order.

There is strong experimental evidence that the antiferromagnetism
and the hidden order in $URu_2Si_2$ are phase-separated and thus
develop independently.\cite{Chandra03}
High-field measurements\cite{Mentink96,vanDijk97,Bourdarot03} 
indicate that the bulk anomalies survive up to  
40 Tesla ($T$), while the staggered moment is destroyed\cite{Mason95}
by comparatively modest fields of 15 $T$. Furthermore,
the staggered magnetic moment grows linearly with pressure\cite{Amitsuka99}
while bulk anomalies associated with the hidden order remain relatively
pressure-independent.\cite{Fisher90} Phase separation is also indicated
by muon spin resonance $(\mu SR)$ experiments.\cite{Luke94,Amitsuka03}
The most direct evidence has come from recent NMR pressure-dependent
measurements (see Figure 2):\cite{Matsuda01} for $T<T_{0}$ the
existence of distinct antiferromagnetic and paramagnetic (hidden order)
phases is clearly observed in samples with less than 
$10\%$ of the volume magnetic
($m_{spin}\approx 0.3\mu _{B}$) at ambient pressure. 
The observed increase of the staggered magnetic moment
with pressure\cite{Amitsuka99} is then simply a 
volume-fraction effect.\cite{Matsuda01}
The magnetic order develops independently from the hidden order through
a first order transition,\cite{Chandra02a} and the associated 
temperature-pressure phase diagram has been determined using thermal
expansion measurements.\cite{Motoyama02}

The mysterious phase transition at $T_0$ has features that have both
local and itinerant electronic natures, and these coexisting dual 
characteristics make its description quite challenging.  For example,
the development of a sharp propagating mode just below $T_0$ observed
by inelastic neutron scattering \cite{Walter86,Broholm87,Mason91}
emphasises the importance of local crystal-field excitations at the
transition. Nevertheless a purely local picture cannot provide a 
straightforward explanation for the observed elastic
anomalies\cite{Luthi93} near $T_0$ 
that are distinct from those of typical uniaxial
antiferromagnets\cite{Melcher70} both due to their (weak) magnitudes  
and due to the absence of precursor effects for $T > T_0$.

\figwidth= 0.4 \textwidth

\fg{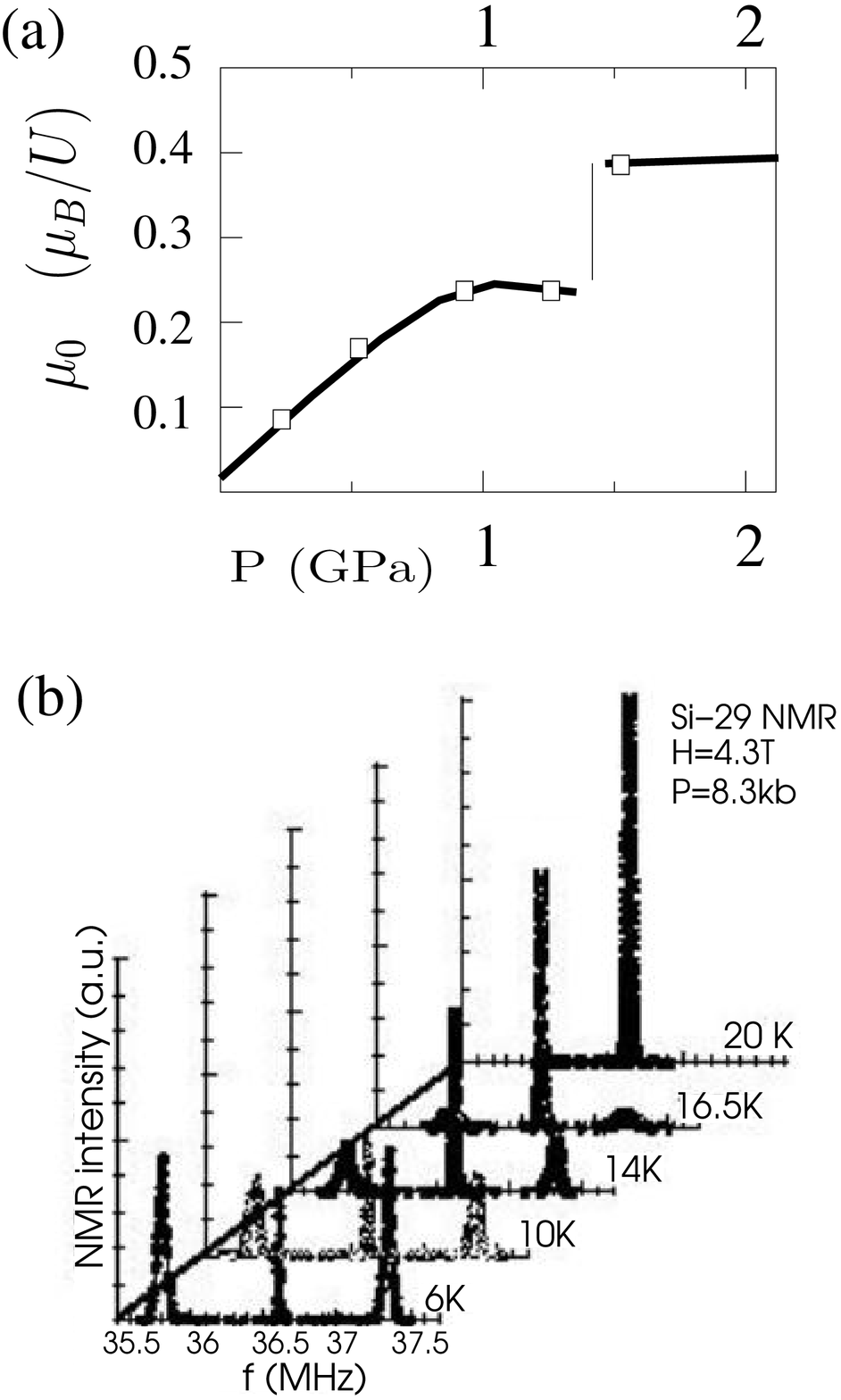}{fig2}{\small Schematic of the (a) pressure
dependence of the ground-state staggered magnetic moment after
ref. [32]. (b)  Satellite structure in NMR after [40],
taken at 0.8GPa, showing
the co-existence of an antiferromagnetic satellite with 
a central peak derived from the hidden order phase. }

The sharp mean-field nature of
the phase transition at $T_0$, together with the magnitude of the
condensation entropy and the observed development of gap in the
excitation spectrum all suggest the development of 
density-wave order within a fluid of itinerant
quasiparticles.\cite{Ramirez92,Chandra94,Chandra02b,Virosztek02}
Itineracy is implicated by the sharpness of the transition while
gap
formation and the large entropy of condensation 
speak in favour of an order parameter at a finite wavevector.
However, a dissenting view on this last point, involving p-wave ferromagnetism,
has recently been proposed.\cite{Varma05}
We note that within the itinerant perspective presented here, there 
are problems matching
details of the excitation spectra as observed in inelastic neutron
scattering experiments.\cite{Broholm91} On the other hand, a purely
local scenario\cite{Broholm91,Santini94} (with anticipated corrections
for itinerant fermions) simply can not be reconciled with the almost complete
quenching of the local moments, implicated by the paramagnetic (as
opposed to Curie-like) susceptibility (see inset Fig. 1b.) and the
large linear specific capacity, normally  associated with well-formed
heavy electrons (Fig. 1a).
There are addition 
inconsistencies with a local picture:  for example, 
the gap $\Delta$ used in the local
singlet scheme\cite{Broholm91} to explain the dispersing magnetic
mode has a {\sl different} field-dependence from that of the bulk $\Delta$
associated with thermodynamic quantities.\cite{Santini00} A strict
adherence to a local scheme requires consideration of many additional
crystal-field levels\cite{Santini00} evolving differently in an applied
field.

A proper theoretical description of the transition at $T_0$ in 
$URu_2Si_2$ must therefore encompass both local and 
itinerant features of the problem.  More specifically, the observed
Fermi liquid properties for $T > T_0$ (e.g. Fig. 1)
combined with the large entropy loss and the sharp nature of
the transition indicate that the underlying quasiparticle
excitations are itinerant, presumably composite objects formed from
the 5f spin
and orbital degrees of freedom of the $U$ ions. Local
physics (e.g. Kondo physics, spin-orbit coupling, crystal-field
schemes)  plays a key role in their development.

We have just outlined a number of general considerations that
we believe are crucial features of the hidden order in $URu_2Si_2$.
Given these criteria, we (with J.A. Mydosh) have proposed that 
it can be described by a general density wave whose form factor is 
constrained by experimental
observation and is ultimately determined by underlying local 
excitations.\cite{Mydosh03} We note that a number of proposals
for the hidden order that fit into this general framework have
been made.\cite{Chandra94,Virosztek02,Amitsuka02,Kiss04,Mineev04,Varma05}
We argue that the large entropy loss at the transition can only
be understood if the density-wave involves the polarisation of a
significant fraction of the quasiparticle band, a condition that discounts
a conventional spin-density wave due to the small size of the observed
magnetic moment.
Taking our cue from ambient-pressure Si NMR measurements (see Figure 3)
that indicate
broken time-reversal symmetry in the hidden ordered phase,\cite{Bernal01}
we (with J.A. Mydosh) have proposed
that $UR_2Si_2$ becomes an incommensurate orbital antiferromagnet at
$T=T_0$ with charge currents circulating between the uranium 
ions.\cite{Chandra02b}  
Here the 
modulation wavevector is chosen to fit the observed isotropic field 
distribution at the silicon sites.  
The resulting real-space fields can then be Fourier
transformed to calculate a neutron scattering structure factor with
a ring of possible q-vectors. Though these results have been presented 
elsewhere,\cite{Chandra02b} in this paper (Section II) 
we provide supporting technical
details and further discussion.  We also determine the NMR linewidths
at the Ru sites. Detailed comparison with recent experiment puts constraints
on the allowed incommensurate wavevectors, allowing us to make more specific
predictions for neutron scattering measurements.

In the second part of this paper, we turn towards an underlying
microscopic picture of orbital antiferromagnetism.  More specifically,
in Section III we explore particle-hole pairings in anisotropic 
incompressible Fermi liquids with specific application 
to $URu_2Si_2.$  Next (Section IV)
we introduce a simple $t-J$ model
with a single heavy band and weak antiferromagnetic spin
fluctuations (AFMSF).  
We note that this particular Hamiltonian was
originally introduced\cite{Miyake86} to describe the AFMSF-mediated transition
in $URu_2Si_2$ at $1.2K$.  We show that this same toy model also supports
particle-hole pairings associated with incommensurate orbital 
antiferromagnetism and quadrupolar charge-density wave formation.  
We end (Section IV) the paper with a summary, and then discuss our results 
in the context
of recent high-field and thermal transport measurements.

\section{Phenomenology and Experimental Predictions}

In this Section we review the experimental motivation for 
incommensurate orbital antiferromagnetism as the hidden order in
$URu_2Si_2$. We develop the phenomenology of this proposal,
independent of microscopic details. The magnitude
and the ordering wavevector of the orbital currents are fitted\cite{Chandra02b}
to the observed isotropic field-distribution at the 
silicon sites as measured by nuclear magnetic resonance (NMR).\cite{Bernal01}
The real-space fields produced 
by the orbital charge currents at all points in the sample volume
are then determined, and we use
this information to make specific predictions for neutron scattering
structure factors and for NMR at non-silicon sites to test this proposal.

\subsection{Incommensurate Orbital Antiferromagnetism as the Hidden Order
in $URu_2Si_2$}

We begin our phenomenological discussion by reviewing the case for
incommensurate orbital antiferromagnetism as the hidden order in $URu_2Si_2$.
There have been many proposals for the primary order parameter in 
this material,\cite{Virosztek02,Amitsuka02,Kiss04,Mineev04,Varma05,Shah00} 
and until recently it was assumed that the spin
antiferromagnetism and the hidden order are coupled and homogenous.
However pressure-dependent NMR measurements,\cite{Matsuda01} supported
by muon spin resonance\cite{Luke94,Amitsuka03} and 
thermal expansion\cite{Motoyama02} data, 
indicate that the hidden and the magnetic orders are 
phase-separated and thus are completely independent.\cite{Chandra03}

\fg{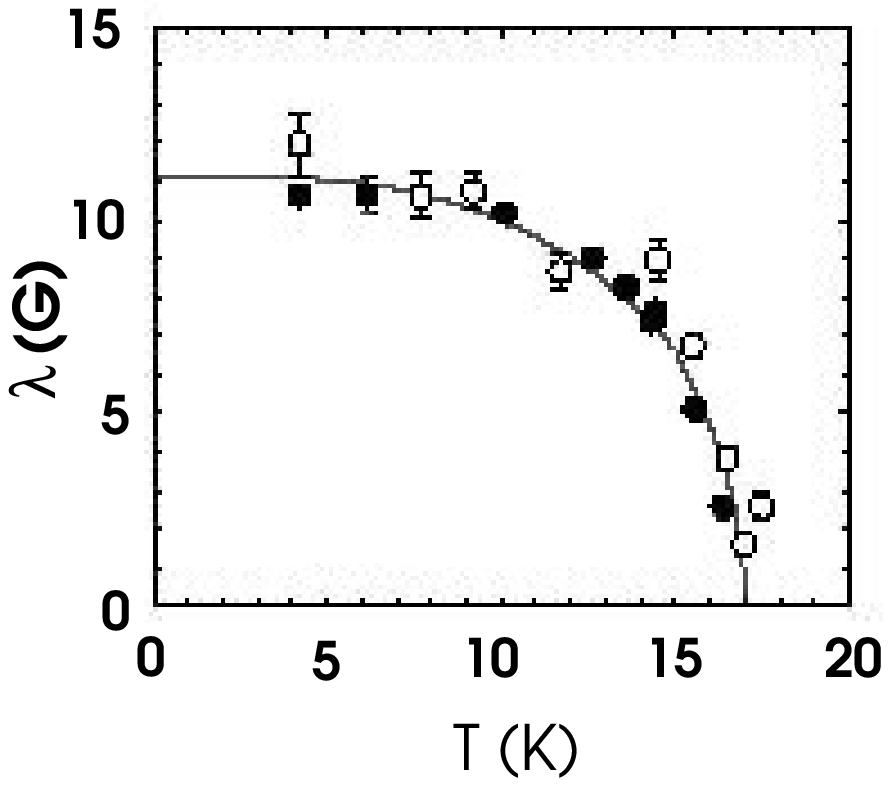}{fig2x}{\small Temperature-dependence of the
NMR line-width $\lambda (T)$, measured in Gauss, 
showing development of a finite local magnetic fields
at $T_{0}$, after [43].
}
We believe that an important clue to the nature of the hidden order 
in $URu_2Si_2$ 
is provided by Si NMR measurements
at ambient pressure\cite{Bernal01} that
indicate that at $T \le T_0$ the paramagnetic (non-split) silicon 
NMR line-width develops a
field-independent, isotropic component whose temperature-dependent
magnitude is proportional to that of the hidden order parameter.
These results imply an isotropic field distribution at the silicon
sites whose root-mean square value is proportional to the hidden order
($\psi$)
\begin{equation}\label{expt}
\langle B^{\alpha } (i) B^{\beta} (j)\rangle = A^{2}\psi ^{2} \delta
_{\alpha \beta }, 
\end{equation}
and is $\sim 10$ Gauss at $T=0$.
This field magnitude is too
small to be explained by the observed moment\cite{Broholm87} that induces
a field
$B_{spin} = \frac{8\pi}{3}\frac{M}{a^3} = 100$ Gauss where
$a$ is the $U-U$ bond length ($a = 4 \times 10^{-8}$ cm).
Furthermore this moment is aligned along the $c$-axis,
and thus cannot account for the isotropic nature
of the local field
distribution detected by NMR.
These measurements indicate that as the hidden order develops,
a static isotropic magnetic field develops
at each silicon site.  This is strong evidence
that the hidden order parameter breaks time-reversal invariance.

The magnetic fields at the silicon nuclei have two possible 
origins:\cite{Schlichter78} the conduction electron-spin interaction
and the orbital shift that is due to current densities.  In $URu_2Si_2$,
the observed Knight shift\cite{Bernal01} indicates a strong Ising
anisotropy of the conduction electron fluid along the $c$ axis; therefore
the electron-spin coupling is unlikely to be responsible for the measured
isotropic field distribution at the Si sites.  It thus seems natural that
these local fields are produced by orbital currents that develop
at $T_0$, and thus we attribute the observed isotropic linewidth to the
orbital shift.

It is this line of reasoning that led us (with J.A. Mydosh) to
propose\cite{Chandra02b} that $URu_2Si_2$ is an incommensurate orbital
antiferromagnet at $T=T_0$ with charge currents circulating between
the uranium ions. The planar tetragonal structure of $URu_2Si_2$
presents a natural setting for an anisotropic charge instability of
this type. We can estimate the local fields at the silicon sites that
are produced by the orbital currents.  On dimensional grounds, the
current along the uranium-uranium bond is given by $I \sim \frac{e
\Delta}{\hbar}$ where $\Delta$ is the gap associated with the
formation of the hidden order at $T_0$; we note that this expression
also emerges from an analysis of the Hubbard model.\cite{Affleck88} If
this orbital charge current is flowing around a uranium plaquette of
side length $a$, then the magnetic field produced at a height $a$
above it is given by Ampere's Law to be $B \approx
\frac{2}{ac}\frac{e\Delta}{\hbar} = 11 G$, in good agreement with the
observed field strength;\cite{Bernal01} here we have used the
experimental value\cite{Palstra85} $\Delta = 110 K$.  Note that the
resulting orbital moment, $m_{OAFM} = 0.02 \mu_B$ ($m_{OAFM} = I
a^2$), is comparable to the effective spin moment at ambient pressure.
We emphasise that an orbital moment produces a field an order of
magnitude less than that associated with a spin moment of the same
value; the low field strengths observed at the silicon sites are
quantitatively consistent with our proposal that they originate from
charge currents.

This orbital moment, $m_{OAFM} = 0.02 \mu_B$, can also account for the
entropy loss at the transition.  We emphasise that its large value
suggests that the amplitude of any proposed density-wave must be
a significant fraction of its maximally allowed value, and will
proceed to show that this is the case for the OAFM.
In a metal the change in the entropy
is given by $\Delta S = \Delta \gamma_n T_0$ where $\Delta \gamma_n$
is the change in the linear specific heat coefficient resulting from the 
gapping of the Fermi surface.  $\Delta \gamma_n$ is inversely proportional
to the Fermi energy $\epsilon_F$ of the gapped Fermi surface, so that
in general the change in entropy per unit cell is given by 
$\Delta {\cal S} \equiv \frac{\Delta S}{k_B} 
\sim (\frac{k_B T_0}{\epsilon_F})$.  
Since the transition
at $T_0$ is mean-field in nature,\cite{Chandra94} we have $\Delta \sim T_0$
so that $\Delta {\cal S} \sim \frac{\Delta}{\epsilon_F}$. Now we recall
that the orbital magnetic moment is
\begin{equation}
m_{OAFM} = I a^2 = 
\left(\frac{e}{\hbar}\right) a^2 \Delta \approx 0.02 \mu_B
\end{equation}
such that it is saturated when $\Delta \sim \epsilon_F$
\begin{equation}
m_{OAFM}^* \sim \left(\frac{e}{\hbar}\right) a^2 \epsilon_F \sim
\left(\frac{a}{a_0}\right)^2 \left(\frac{\epsilon_F}{\epsilon_H}\right) \mu_B
\sim 0.1 \mu_B
\end{equation}
analogous to the saturation value of the electron spin
$\mu_B = \left(\frac{e}{\hbar}\right) a_0^2 \epsilon_H$
where $a_0$ and $\epsilon_H$ are the Bohr radius and the energy
of the Hydrogen atom respectively; here we have used
$\frac{a}{a_0} \sim 10^2$ and 
$\frac{\epsilon_F}{\epsilon_H} \sim \frac {M_H}{M^*} \sim 10^{-3}$
where $M_H$ and $M^*$ refer to the mass of hydrogen and of $URu_2Si_2$
respectively. 
Then the change in entropy
at the transition 
($\Delta {\cal S} \sim \frac{\Delta}{\epsilon_F}$)
due to the development of orbital
antiferromagnetism can be expressed as
\begin{equation}
\Delta {\cal S}_{OAFM} \approx \left(\frac{m_{OAFM}}{m_{OAFM}^*}\right) 
\approx 
0.02 \left(\frac{\mu_B}{m_{OAFM}^*}\right) \approx 0.2 
\end{equation}
which is a number ($0.2 = 0.3 \ln 2$) in good agreement with 
experiment.\cite{Palstra85}
We also note that the critical field for suppressing the thermodynamic
anomalies is distinct from its spin counterpart: the ratio 
$\frac{H_c^{orb}}{H_c^{spin}} \sim  \frac{\mu_b}{m_{OAFM}^*} \sim 10$ 
is qualitatively
consistent with the observed critical field associated with the destruction
of hidden order.\cite{Mentink96,vanDijk97}  We emphasise that the sizable
entropy loss associated with the development of orbital antiferromagnetism
in $URu_2Si_2$ is a direct consequence of its renormalised electron
mass ($\frac {M^*}{M} \propto \frac{\epsilon_H}{\epsilon_F}$). 
More generally the orbital
moment is a larger fraction of its saturation value than is its spin 
counterpart, and this leads to the large entropy loss.
     
Orbital antiferromagnetism can therefore account for the local field
magnitudes at the silicon ions and for the large
entropy loss at the transition.  Our next step is to tune the ordering
wavevector to fit the isotropic distribution at these sites and then
to determine the real-space fields throughout the sample volume.  
This can then be Fourier transformed
to make predictions for neutron scattering.\cite{Chandra02b}  
We note that it has been suggested\cite{Varma05} that
the isotropic nature of the field distributions at the silicon sites
may be due to impurity-broadening. 
Though disorder is certainly present in these samples,
we believe that the
incommensurate nature of the density wave is the origin of this
isotropy.  Towards proving this point, we have determined the
anisotropic field distributions at non-silicon sites; their observation
via NMR would certainly not be possible if there were significant 
disorder-smearing.
  
Before proceeding with this program, let us comment briefly on
the current experimental situation regarding the proposal of incommensurate
orbital antiferromagnetism in $URu_2Si_2$.
We admit that our proposal is closely linked to the ambient-pressure NMR
experiments,\cite{Bernal01} which are the only direct evidence of
broken time-reversal symmetry in the hidden ordered phase and have
not been reproduced by other groups.
We note that muon spin resonance measurements\cite{Luke94,Amitsuka03}
support the emergence of local fields with the same temperature-dependence
as that associated with NMR, but their overall amplitudes are two 
orders of magnitude
less than that seen in the NMR measurements. 
This is a point to which we return in the discussion. 
Although incommensurate peaks have been seen in inelastic neutron
scattering measurements,\cite{Broholm91,Bull02,Wiebe04},  these
are due to excitations above the partly gapped Fermi surface and are
not directly related to the orbital antiferromagnetism.
Current experimental
resolution for elastic scattering - a direct probe of the
incommensurate orbital antiferromagnetic order - 
is not yet good enough to confirm or deny the OAFM scenario. 
Here we present technical support for previous predictions for 
neutron structure factors,\cite{Chandra02b} while also making
specific suggestions for measurements where the signal should
be sufficiently strong to be observed practically.

\subsection{Predictions for Neutron Scattering}

In order to calculate the neutron cross section for scattering
by incommensurate orbital antiferromagnetic order, we use the
Born scattering formula,
\begin{equation}
\frac{d\sigma}{d\Omega}=\left(\frac{g_{N}e^{2}}{8\pi\hbar c}\right)^{2}
|{\bf B}({\bf q})|^2 = r_{0}^2 S({\bf q}),
\label{crossection1}
\end{equation}
where $g_{N}$ is the neutron gyromagnetic ratio, ${\bf q}$ the
scattering wavevector of the neutrons, 
$|{\bf B}({\bf q})|^2$ is the structure factor of the magnetic  
fields produced by the orbital currents and 
$S({\bf q})=|{\bf B}({\bf q})|^2/(4\pi\mu_{B})^{2}$ is the structure factor measured in units
of the Bohr magneton $(\mu_{B})$. 

We shall compute the magnetic field as the curl of the vector
potential, ${\bf B}({\bf x})={\bf \nabla} \times {\bf A}$.   The
procedure will be to compute the vector potential 
produced by the circulating current 
around a given plaquette.
We shall
denote the co-ordinate of the centre of plaquette j by ${\bf{X}}_{j}$. 
The corners of this plaquette are located at sites ${\bf{x}}_{j}^{(r)}$,
($r=1,4$) where 
\[
{\bf{x}}_{j}^{(r)} = {\bf{X}}_{j}+{\bf{x}}^{(r)}, \qquad  (r=1,4), 
\]
as shown in Fig. \ref{links1} (a).
The
circulating current around plaquette $j$ is then taken to be
\begin{equation}\label{l1}
I_{C}{(\bf{ X}_{j})} = I_{0} e^{i \bQ \cdot {\bf{ X}_{j}}} + {\rm H.c}
\end{equation}
Using Ampere's law, link 1-2 will 
will produce a contribution 
to the vector potential given by
\begin{equation}
{\bf A}^{12}({\bf x})=\frac{1}{c}\sum_{j}\int_{{\bf x}_{j}^{(1)}}^{{\bf
    x}_{j}^{(2)}}
dx' \frac{I_{C}
({\bf X}_{j})\hat {\bf x}_{12}
}{|{\bf x}-{\bf x'}_{j}|},
\label{ampere} 
\end{equation}
where $\hat {\bf x}_{12}
$ is the unit
vector pointing along the bond from $1$ to $2$.
Writing ${\bf x'}_{j}$ as 
\[{\bf x'}_{j}=
 {\bf
  x}_{j}^{(1)} 
+ w\,({\bf x}^{(2)} - {\bf x}^{(1)}),\]
where $0 < w < 1$ defines the position along the link,
we have 
\begin{equation}
{\bf A}^{12}({\bf x})=\frac{a  }{c}
\sum_{j}\int_{0}^{1} 
dw\,
\frac{{ I}_{C}({\bf X}_{j})\ \hat {\bf x}_{12}
}{|{\bf x}- \{ {\bf
    x}_{j}^{(1)} + w({\bf x}^{(2)}-{\bf x}^{(1)})\}|}.
\label{A}
\end{equation}
 for the vector potential where $a$ is the U-U bond length in the $ab$ plane.

\figwidth= 0.5 \textwidth

\fg{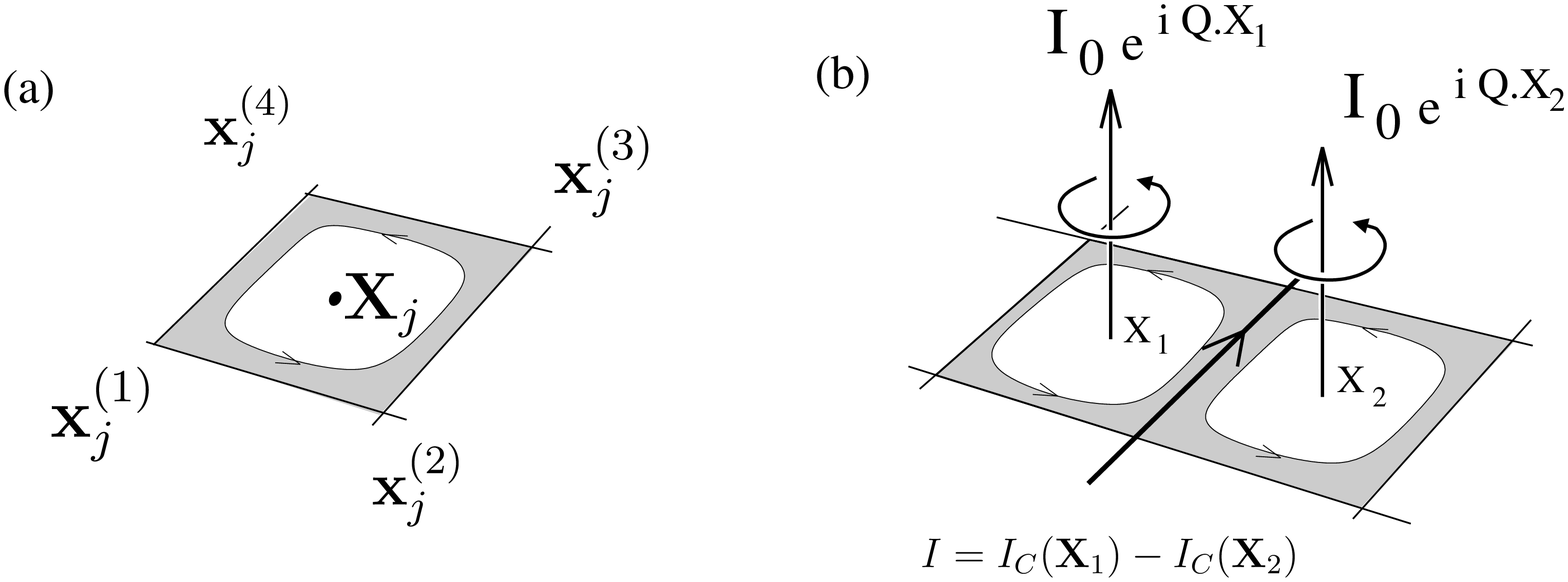}{links1}
{\small (a) Labelling of
sites around a single plaquette.
(b) Schematic of circulating currents
    $I_{C} ({\bf X}_{j}) =I_{0}\exp[i\mathbf{Q}\cdot\mathbf{x}_{j}]$ flowing in the uranium
    plaquettes in the $ab$ plane. The plaquettes are labelled by the
    coordinates of their centre ${\bf{X}}_{j}$. 
}

\begin{widetext}
We now compute ${\bf{B}}^{12}=\nabla \times {\bf{A}}^{12}$, and take
the Fourier transform to obtain 
\begin{eqnarray}
{\bf B}^{12}({\bf q}) & = & 
\frac{a}{c}\sum_{j}\int_{0}^{1} dw\,\int d^{3}{ x}\, 
e^{-i{\bf q}.{\bf x}}\, 
{ I}_{C}({\bf X}_{j})\ \hat {\bf x}_{12} \times {\bf \nabla}
\frac{1}{|{\bf x}-\{{\bf X}_{j}+ {\bf
    x}^{(1)} + w({\bf x}^{(2)}-{\bf x}^{(1)})\}|}
\nonumber \\
 & = & \frac{ia}{c}\sum_{j} 
{ I}_{C}({\bf X}_{j})\ \hat {\bf x}_{12} \times {\bf
   q} \int_{0}^{1} dw\,\int d^{3}{x}\, e^{-i{\bf q}.{\bf x}}\, 
\frac{1}{|{\bf x}-\{{\bf x}_{j}+ {\bf
    x}^{(1)} + w({\bf x}^{(2)}-{\bf x}^{(1)})\}|}.
\end{eqnarray}
Using
\[
\int d^{3}e^{-i \bq \cdot\bx} \frac{1}{\vert \bx  - {\bf{a}}\vert} =
\frac{4\pi}{q^{2}}e^{-i\bq \cdot{\bf{a}}
},
\]
we obtain
\begin{eqnarray}\label{}
{\bf B}^{12}({\bf q}) & = & 
  \frac{i4\pi a}{q^{2} c}\sum_{j} 
{ I}_{C}({\bf   X}_{j}) \hat {\bf{x}}_{12}
\times {\bf q} \int_{0}^{1} dw\, \exp[-i{\bf q}.({\bf x}_{j}+{\bf
  x}^{(1)} + w({\bf x}^{(2)}-{\bf x}^{(1)}))]
\nonumber \\
 & = &  -\frac{4\pi a}{q^{2} c}\sum_{j} 
e^{-i{\bf q}.{\bf X}_{j}} { I}_{C}({\bf   X}_{j})\frac{
 \hat {\bf{x}}_{12}
 \times {\bf q}}{{\bf q}.({\bf x}^{(2)}-{\bf
   x}^{(1)})}
(e^{-i{\bf q}.{\bf x}^{(2)}}-e^{-i{\bf q}.{\bf x}^{(1)}})\cr
&\equiv & \frac{4\pi I_{0}}{q^{2} c}
\left({\bf F}^{12}({\bf q}) \times {\bf q}  \right)
\sum_{j}
e^{i{(\bQ -\bf q)}.{\bf X}_{j}} 
,
\label{Bq}
\end{eqnarray}
where we have replaced $I_{C} ({\bf{X}_{j}})= I_{0}e^{i \bQ \cdot {\bf{X}}_{j}}
$ and 
\begin{equation}\label{}
{\bf F}^{12}({\bf q})= \frac{\hat {\bx}_{12}}{\bq \cdot
\hat \bx_{12}}
(e^{-i \bq \cdot \bx^{(1)}}-e^{-i \bq \cdot \bx^{(2)}})
\end{equation}
is the form-factor associated with link $1-2$
in the plaquette centred about ${\bf X}_{j}$.
To sum over all of the links around the plaquette, we must add together
the form factors
\begin{eqnarray}
{\bf{ F}} (\bq )&=& \label{F}
{\bf{ F}}^{12} (\bq )+
{\bf{ F}}^{23} (\bq )+
{\bf{ F}}^{34} (\bq )+
{\bf{ F}}^{41} (\bq )\cr
&=&\bigg[
\frac{{\hat{\bf x}}}{{\bf q}.{\hat {\bf x}}}\{
e^{i{\bf q}.({\hat
      {\bf x}}+{\hat {\bf y}})a/2} 
- e^{-i{\bf q}.({\hat
      {\bf x}}-{\hat {\bf y}})a/2} 
\nonumber \\
 &   & \qquad\qquad\qquad + e^{-i{\bf q}.(-{\hat
      {\bf x}}+{\hat {\bf y}})a/2}
- e^{-i{\bf q}.({\hat
      {\bf x}}+{\hat {\bf y}})a/2}
\}-{\hat {\bf x}}
\leftrightarrow {\hat {\bf y}}\bigg] \nonumber \\
 & = & 4\sin\left(\frac{q_{x}a}{2}\right)\,
       \sin\left(\frac{q_{y}a}{2}\right)
\left\{    \frac{{\hat {\bf y}}}{{\bf q}.{\hat {\bf y}}}
-
  \frac{{\hat {\bf x}}}{{\bf q}.{\hat {\bf x}}} 
\right\}. 
\end{eqnarray}

We let ${\bf Q}$ be the wavevector for the orbital order so that
$I({\bf x}_{j}) = I_{0}\exp[-i{\bf Q}.{\bf x}_{j}]$.

Replacing ${\bf{F}}^{12}\rightarrow {\bf{F}}$ in Eq. \ref{Bq}, we 
we obtain the complete
Fourier transform of the magnetic field:
\begin{equation}\label{Bq2}
{\bf B}({\bf q}) = 
\sum_{j}
\exp[i({\bf Q}-\bq).{\bf x}_{j}]
\sin\left(\frac{q_{x}a}{2}\right)
       \sin\left(\frac{q_{y}a}{2}\right)
\left\{ 
\frac{\hat {\bf y}}
{{\bf q}\cdot {\hat {\bf y}}} -  
   \frac{\hat {\bf x}}{{\bf q}.{\hat {\bf x}}}
\right\}
\times {\bf  q}.
\end{equation}
\end{widetext}

The U sites ${\bf x}_{j}$ can be written as 
\begin{equation}
{\bf x}_{j} = a\,(j_{1},j_{2},0) + \frac{c}{2}\,(0,0,j_{3})
                 +
                 \frac{1}{2}(1-(-1)^{j_{3}})\,(\frac{a}{2},\frac{a}{2},0),
\label{Usites}
\end{equation}
where $c$ is the separation between even or odd numbered U planes. The  
unit cell has lattice vectors $(a,0,0),(0,a,0),(0,0,c)$.
For an isotropic distribution of magnetic fields at the Si sites, we
can reasonably expect ${\bf Q}$ to be staggered between successive U
layers. We permit ${\bf Q}$ to be
incommensurate in the $a-b$ plane:
\begin{equation}
{\bf Q} = (Q_{x},Q_{y},0) + \frac{2\pi}{c}\,(0,0,1).
\label{Q}
\end{equation}
Summing over the lattice sites in Eq.(\ref{Bq2}) we find

\begin{eqnarray}
{\bf B_{OAFM}}({\bf q}) & = &  \frac{8\pi I_{0}}{q^{2}c}
\sum_{{\bf G}}\delta_{{\bf q},{\bf Q}+{\bf G}}\left[1+e^{i{\bf G}\cdot
 (a/2,a/2,c/2)}\right]\times \nonumber \\
 & \times & \!\!\!\sin\left(\frac{q_{x}a}{2}\right)\,
       \sin\left(\frac{q_{y}a}{2}\right)
\left\{ \frac{{\hat {\bf x}}}{{\bf q}.{\hat {\bf x}}} -  
   \frac{{\hat {\bf y}}}{{\bf q}.{\hat {\bf y}}}\right\}\times {\bf
 q},
\label{Bq3}
\end{eqnarray}
where ${\bf G}=2\pi[n_{1}/a,n_{2}/a,n_{3}/c]$ is a reciprocal lattice vector.

Equation (\ref{Bq3}) should be contrasted with the corresponding expression
if the order parameter were a spin density wave instead of an orbital 
antiferromagnet:
\begin{eqnarray}
{\bf B}_{SDW}({\bf q}) & = &  \frac{4\pi}{c}
\sum_{{\bf G}}\delta_{{\bf q},{\bf Q}+{\bf G}}\left[1+e^{i{\bf G}\cdot
 (a/2,a/2,c/2)}\right]\times \nonumber \\
  & & \qquad\qquad\qquad \times \left\{{\hat {\bf q}}\times ({\bf M}\times {\hat {\bf q}})\right\}.
\label{Bq3SDW}
\end{eqnarray}
We note that a major difference between the two cases
is that ${\bf B_{OAFM}}({\bf q})$ decreases
rapidly as $q^{-2}$ while ${\bf B}_{SDW}({\bf q})$ is
constant. This makes OAFM much harder to detect in neutron scattering
experiments than its SDW counterpart.
Second, the term 
$({\bf q}\times ({\bf M}\times {\bf q}))$ in   
${\bf B}_{OAFM}(q)$ indicates that scattering is suppressed for
${\bf q}={\bf Q}$ since for an SDW along
the $c$ axis, ${\bf M}\parallel {\bf Q}=\frac{2\pi}{c}(0,0,1)$. 
here is no such term in ${\bf B}_{OAFM}(q)$.
Thus the presence of a finite scattering amplitude at this particular
wavevector in $URu_2Si_2$ would be a ``smoking gun'' confirmation
of incommensurate orbital antiferromagnetism as the hidden order.

Next we turn to obtaining the structure factor $|{\bf B}({\bf q})|^2$. 
Neutrons couple to the orbital currents via their 
magnetic moment (${\pmb \mu}_{N}=g_{N}\mu_{B}{\bf S}$) as 
$E=-{\pmb \mu}_{N}.{\bf B}$. For incoherent neutrons, $|{\bf B}({\bf q})|^2$
is the modulus squared of ${\bf Q}$ averaged over the orientation.
the neutrons. Thus
\begin{widetext} 
\begin{eqnarray}
S({\bf q}) & = & \frac{|{\bf B}({\bf q})|^2}{(4\pi\mu_{B})^2} 
  = \left(\frac{NI_{0}a^2}{c\mu_{B}}\right)^2
\sum_{{\bf G}_{n_1,n_2,n_3}}\delta_{{\bf q},{\bf Q}+{\bf G}}\,
\left\{j_{0}\left[\frac{q_{x}a}{2}\right]\,j_{0}\left[\frac{q_{y}a}{2}\right]
\right\}^2 \times \nonumber \\
 &   & \qquad\qquad\qquad
  \times\left[\frac{1+\cos[\pi(n_{1}+n_{2}+n_{3})]}
{2}\right]^2\,
\frac{q_{x}^2+q_{y}^{2}}{q_{x}^2+q_{y}^2+q_{z}^2},
\label{Sq2}
\end{eqnarray}
\end{widetext}
where $j_{0}(x)=\frac{\sin x}{x}$ and $N$ is the number of U sites.
From this expression, we find that the maximum scattering intensity
is predicted\cite{Chandra02b} to lie in a ring $\vec{Q} = \vec{Q_0} + \vec{q}$ of radius $|q| \sim 0.2$ centred on the wavevector $\vec{Q_0} = (001)$
where $\vec{q}$ lies in the $a-b$ plane.  Once again, we emphasise that scattering
in the vicinity of $\vec{Q_0}$ is {\sl forbidden} for the case of
ordered spins along the $c$-axis; thus the observed presence
of neutron scattering intensity at this particular wavevector would
be a ``smoking gun'' confirmation of orbital antiferromagnetism as
the hidden order.

In general the structure factor can be written as a product
\begin{equation}
S(q) = f(q) g(q)
\end{equation}
where $g(q)$ is a function periodic in the reciprocal lattice vector
but $f(q)$ is not.  For the case of orbital antiferromagnetism, the
calculated structure factor yields an asymptotic
form for the form factor $f(q) \sim \frac{1}{q^4}$. This power-law
decay of the intensity peaks is due to the extended nature of the
scattering source in contrast to the exponentially decaying structures
observed for point-like spin antiferromagnetism.

It is tempting to state that such power-law peaks will
be a clear signature of orbiting charge currents, but we still
need to determine whether the overall intensities are observable.
We can estimate the strength of the predicted OAFM neutron signal
compared to that associated with spin magnetism at ambient pressure.
Our calculations indicate that a fifth of the total integrated weight
of $S(q)$ (TIWSQ) resides in the first Brillouin zone for the OAFM.
Using the sum rule that relates the total ISWQ (integrated weight of
$S(q)$ to the square of
the moment, we have
\begin{widetext} 
\begin{eqnarray}
(IWSQ)_{BZ1} & = & \frac{1}{5} (TIWSQ)_{OAFM} = 
\frac{1}{5} (m_{OAFM})^2  \\
& = & \frac{1}{500} (m_{spin})^2 = \frac{1}{500} (TIWSQ)_{spin}
\end{eqnarray}
\end{widetext}
where we have used $m_{OAFM}=0.2\mu_B$ and $m_{spin}=0.3 \mu_B$.
Since the magnetic region occupies roughly a tenth of the sample at
ambient pressure we then write
\begin{equation}
(IWSQ)_{BZ1} = \frac{1}{50} \quad {\rm Measured} \\  (TIWSQ)_{spin}
\end{equation}
which indicates that the scattering peaks in the first Brillouin zone
due to orbital ordering should have roughly 1/50 the intensity of the
analogous spin peaks at ambient pressure.
There have been two exploratory neutron 
studies\cite{Bull02,Wiebe04}
but neither was conclusive due to issues of resolution.  In particular
the more recent elastic measurements\cite{Wiebe04} were not
performed at the predicted wavevector
$\vec{Q_p} = (\tau_p,\tau_p,1)$ where there should be no dipole
scattering; please recall that here the form factor 
$\sim (\vec{q} \times \vec{m}$)
and $\vec{m}$ is aligned with the c-axis.  More specifically
the scattering intensity
should be a factor of twenty higher than at 
$\vec{Q_e} = (1 + \tau_x,\tau_y,0)$ where the experiments were performed,
and the experimental resolution should be good enough then to prove/refute
the orbital antiferromagnetism proposal.

\subsection{Nuclear Magnetic Resonance Linewidth at the Si and the Ru Sites}

Nuclear Magnetic Resonance (NMR) is a local probe of the strength and the local
distribution of the magnetic field distribution in the material.   
We use experimental NMR results to determine the 
ordering wavevector associated with the orbital antiferromagnetism,
which can then be included in the structure factor calculated above.
Thus neutron scattering and NMR are 
complementary. Eq.(\ref{A}) gives the vector potential at a point
${\bf x}$ due to a current in link $\langle 12\rangle$ of a 
plaquette centred at ${\bf X}_{j}$. Contributions from other 
links in the plaquette may be
similarly written out (see Fig.\ref{links1}). The magnetic field at
any point ${\bf x}$ can be obtained using ${\bf B}={\bf \nabla}\times
{\bf A}$, where ${\bf A}$ is the total vector potential obtained by
summing contributions from all links and
plaquettes. We give detailed expressions for
${\bf A}$ in the appendix.  

For the sake of completeness, we review\cite{Chandra02a,Chandra02b} 
our arguments regarding
the Si NMR measurements\cite{Bernal01} and the ordering wavevector
of the orbital antiferromagnetism. We note that the silicon atoms
in $URu_2Si_2$ are located at low-symmetry sites above and below
the uranium plaquettes, so that the fields there do not cancel. Therefore
the proposed OAFM must have an incommensurate ${\bf Q} \neq (\pi,\pi)$
in order to produce isotropic field distributions at the silicon sites.
If the order parameter in the hidden order phase is OAFM, then 
such a magnetic field distribution at the Si sites would
be possible if the wavevector for orbital ordering were 
incommensurate,\cite{Chandra02a,Chandra02b}
\begin{equation}
{\bf Q} = \frac{2\pi}{a}(0.22\cos\phi,0.22\sin\phi,a/c).
\label{incomm}
\end{equation}
Fig.(\ref{f:magfield}) shows the distribution of
the magnetic field lines about the $ab$ plane for an incommensurate $\mathbf
Q$ corresponding to $\phi =\pi/4$ in Eq.(\ref{incomm}), and viewed in the
[010] direction.
\fg{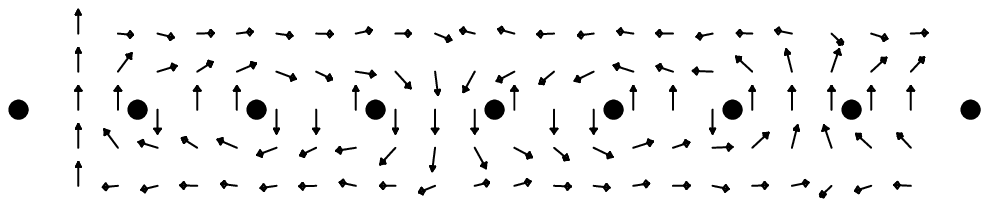}{f:magfield}
{\small The distribution of
the magnetic field lines about the $ab$ plane for an incommensurate ${\mathbf
Q}=\frac{2\pi}{a}(0.16, 0.16, a/c),$ and viewed in the
[010] direction. The black circles represent Uranium atoms in the $ab$ plane.}

A convenient definition of the anisotropy in the magnetic field
at a given site is
\[
\zeta=|(B_{\perp}-B_{\parallel})/(B_{\perp}+B_{\parallel})|, 
\]
Fig.(\ref{sifield}) shows the anisotropy as a function of the
$\bf{ Q}$ vector.
While the field distribution at the Si sites
is isotropic, that need not be the case at other sites such
as Ru; furthermore the {\sl anisotropic} nature of the field
distribution at the Ru sites would indicate that disorder-averaging
is not at play here. 
If we take as the origin any uranium atom in the lattice, we find the Ru sites
at coordinates
\begin{equation}
{\bf X}_{\mbox{Ru}} = \frac{a}{2}(i-j+1,i+j,0) +
\frac{c}{2}(0,0,k+1/2),
\label{Rucoods}
\end{equation}
where $i,j,k$ are integers. 
Fig.\ref{rufield} shows the anisotropy
of
the magnetic field distribution at the Ru sites.  
\figwidth = 0.5\textwidth
\fg{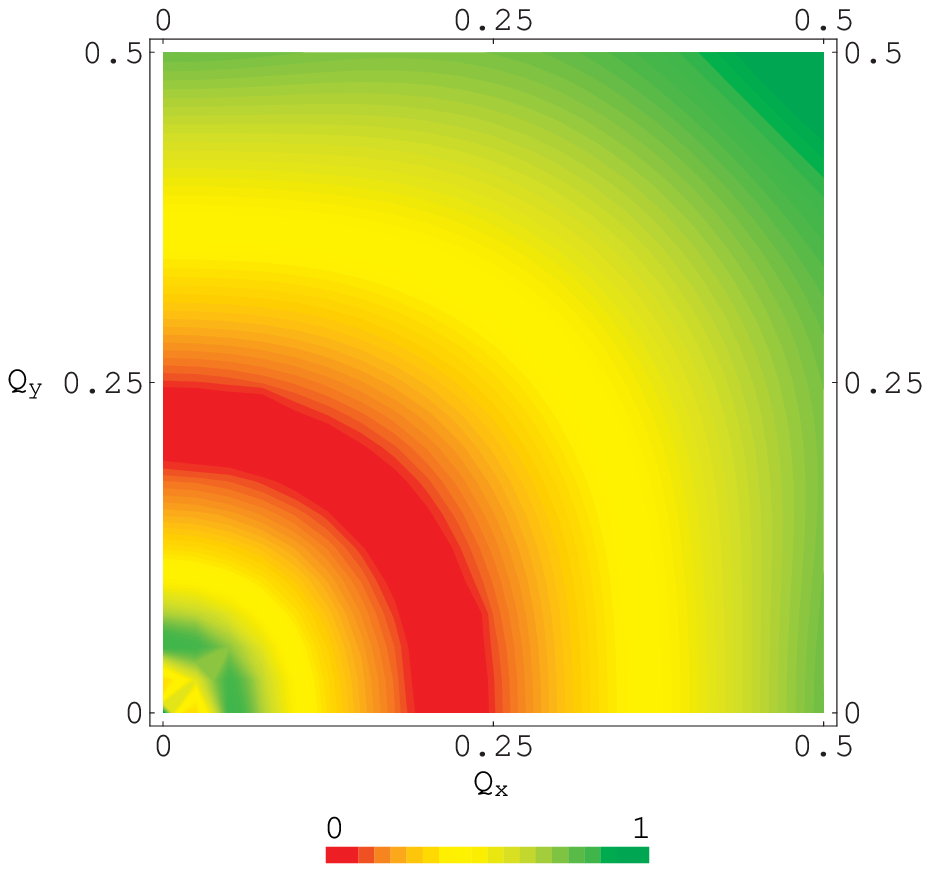}{sifield}
{\small Anisotropy of magnetic field distribution at the Si
    sites. The anisotropy is found to vanish on a ring of wavevectors
    approximately given by 
${\bf Q}=
\frac{2\pi}{a}
(0.22\cos\phi,0.22\sin\phi,a/c)$.}

\fg{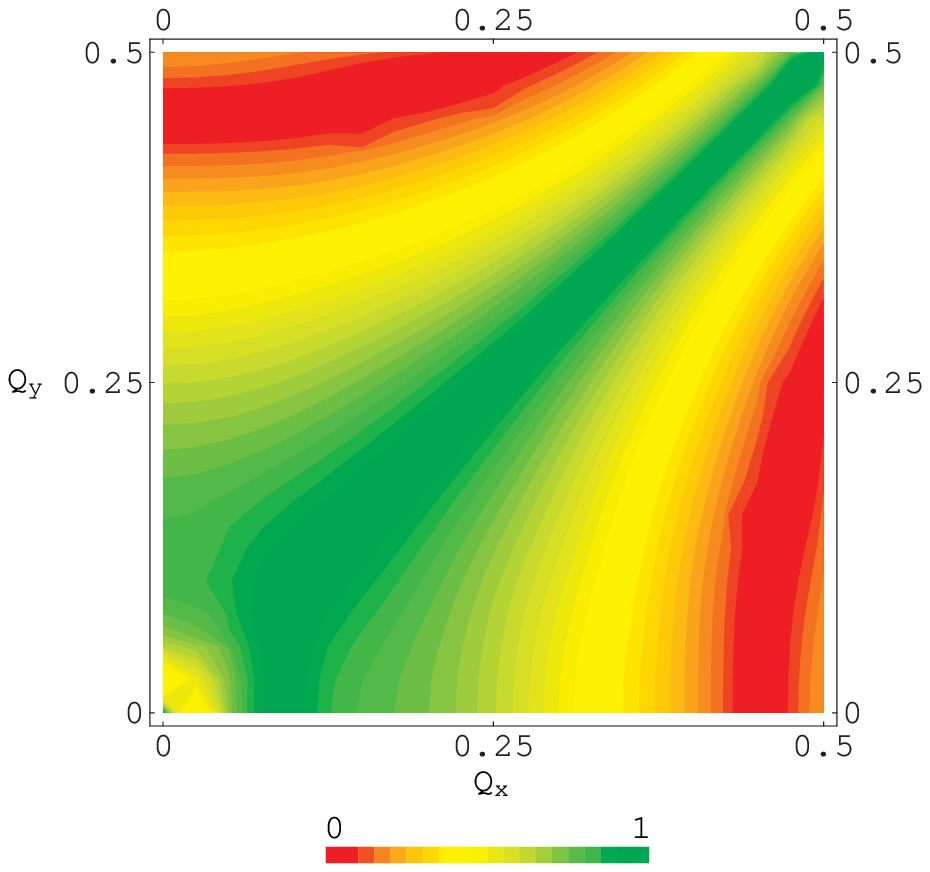}{rufield}
{{\small The anisotropy of magnetic field distribution at the
    Ru sites for orbital wavevector of the form ${\bf Q}=
    (Q_{x},Q_{y},2\pi/c)$. Darker shades indicate lower anisotropy.}}

Recent Ru NMR measurements\cite{Bernal04} report a local magnetic 
field anisotropy of around 0.3. Values of
$\mathbf{Q}$ deduced from our OAFM model using the Ru NMR data should
of course be consistent with Si NMR. The anisotropy of the magnetic field
at the Ru sites calculated from our model shows strong variations as
the orientation of the incommensurate wavevector given in
Eq.(\ref{incomm}) is varied. Anisotropy of field at the Ru sites for
OAFM ordering wavevectors given by Eq.(\ref{incomm}) varies from about
0.7 along the $\phi=0,\pi/2$ 
directions to nearly unity along $\phi=\pi/4$. 
Thus the most likely
incommensurate wavevector $\mathbf{Q}$ for OAFM lies close to the
$\phi=0,\pi,\pm \pi/2$ directions. 

Neutron scattering measurements\cite{Wiebe04}
show enhanced scattering for $ T> T_{0}$ at the incommensurate wavevectors 
\begin{equation}
\mathbf{Q}_{exp}= (2\pi/a)(n_1+0.4\cos\phi, n_2+0.4\cos\phi,n_3),
\label{Qexp}
\end{equation}
where $n_{1}+n_{2}+n_{3}$ is an odd integer. Below $T_{0}$, the ring
of excitations seems to collapse toward the $x$ and $y$ directions,
decreasing in intensity. The structure-factor predicted in Eq.(\ref{Sq2})
could not be verified/refuted due to issues of resolution.\cite{Wiebe04}
According to Eq.(\ref{Sq2}), the structure factor 
measured near $\mathbf{Q}_{exp} = (2\pi/a)(1.4,0,0)$, as was done
in the most recent experiment\cite{Wiebe04}
has a scattering intensity that is 
smaller than that at $\mathbf{Q} = (2\pi/a)(0.4,0,a/c)$ 
by a factor of more than five. 
In an earlier experiment,\cite{Broholm91} enhanced scattering was observed at
$\mathbf{Q}=(2\pi/a)(1.4,0,0)$ above the transition temperature
$T_{0}$. The scattering
intensity was sharply enhanced for $T<T_{0}$, and
furthermore, the scattering linewidth decreased to resolution-limited
values. More work is needed to verify whether the incommensurate peak
observed in neutron scattering measurements is related to
Eq.(\ref{incomm}) deduced from Si and Ru NMR data using our model of
orbital antiferromagnetism, and we strongly suggest elastic neutron
scattering measurements at the wavevector predicted to have the
greatest intensity ($\mathbf{Q} = (2\pi/a)(0.4,0,a/c)$) to test
OAFM as hidden order. 

\begin{table*}

\begin{tabular}{|c|c|c|c|}
\hline 
Name&
$A_{\mathbf{k}}^{\sigma \sigma '}(\mathbf{Q})$&
 T-invariance&
 Local Fields\\
\hline
SDW (isotropic spin density wave)&
 $\sigma $&
 no&
 yes\\
\hline
CDW (isotropic charge density wave)&
 const.&
 yes&
 no\\
\hline
d-SDW&
 $\sigma \, (\cos (k_{x}a)-\cos (k_{y}a))$&
 no&
 no\\
\hline
q-CDW (quadrupolar)&
 $\cos (k_{x}a)-\cos (k_{y}a)$&
 yes&
 no\\
\hline
OAFM (orbital antiferromagnet)&
 $i\, (\sin (k_{x}a)-\sin (k_{y}a))$&
 no&
 yes \\
\hline
\end{tabular}

\caption{Possible symmetries for particle-hole pairing}
\label{symmetries}
\end{table*}

\section{Towards a Microscopic Description of the Hidden Order}

We now turn to a more microscopic approach to the hidden order.
As we have already noted, a proper theoretical description of 
$URu_{2}Si_2$ must encompass
both local and itinerant features of the problem. A general duality scheme
for heavy electron systems has been
proposed.\cite{Kuramoto90}
In this model, the itinerant excitations are constructed
from the low-lying crystal-field multiplets of the uranium atom. The
quasiparticles associated with the heavy Fermi liquid are composite
objects formed from the localised orbital and spin degrees of freedom
of the U ions and the conduction electron fields. The phase transition
in this model is then a Fermi-surface instability of these composite
itinerant f-electrons.
This approach has been adapted\cite{Okuno98}
to describe the coexistence of hidden order with a small moment
in $URu_2Si_2$.  With the more recent understanding that the hidden
ordered phase does not contain a staggered magnetisation, we have revisited
this duality scheme\cite{Mydosh03} and, guided by experiment, now discuss
its implications for the nature of the mysterious order that develops
at $T = T_0$.

\subsection{Possible symmetries for particle-hole pairing}

We begin with the assumption that all the excitations of
$URu_{2}Si_{2}$ that condense into the hidden ordered state are of
itinerant character.  More specifically, we will assume that 
all of the system's local physics
(e.g. local moment character of the f-electrons) 
has been absorbed into the formation of composite quasiparticles. 
Given this premise, it then follows that key aspects of the
(hidden) order parameter will be expressed through its matrix elements between 
quasiparticle states. If we denote it by the  
operator $\hat \Psi $, then its general matrix element  
between quasiparticle states is
\begin{equation}
\langle {\bf{k}}+{\bf{Q}/2},\ 
\si\vert  \hat \Psi \vert
{\bf{k}}-{ \bf{Q}/2},\si'\rangle  = A_{\bk}^{\si \si'}(\bf{Q})
\end{equation}
where $\mathbf{Q}$ is the ordering wave-vector and $\vert
{\bf{k}}\si\rangle$ is the quasiparticle state of momentum $\bk $. 
Microscopically we would have to characterise $\hat  \Psi $ 
in terms of the detailed crystal-field split states of the U ion, but 
for the purposes of characterising the phase transition, 
quasiparticle matrix elements should suffice. Within the Hilbert space
of the mobile f-electrons, the order parameter can then be written
\begin{equation}\label{phpairX}
\hat \Psi \equiv A_{\bk }^{\si \si'}(\bf{Q})c\dg_{\bk +{\bf Q}/2, \ \si}c_{\bk
- {\bf Q}/2,\ \si'}.
\end{equation}
where $A_{\bk}^{\si \si'} (\bf{Q})$ is a general function of spin and 
momentum.

We are therefore considering a class of density waves with the most 
general pairing in the particle-hole channel characterised by 
$A_{\bk}^{\si \si'}(\bf{Q})$. 
We now categorise the possible particle-hole
pairings\cite{Mydosh03} in $URu_2Si_2$.  Assuming 
that the hidden order develops
between the uranium atoms in each basal plane, we restrict
our attention to nearest-neighbour pairings on a two-dimensional 
square lattice, and display the five resulting possibilities in
Table \ref{symmetries} in Eq.(\ref{phpairX}).   
We emphasise that each of these pairing 
choices will partially
gap the Fermi surface, accounting for the large entropy loss and the observed
anomalies in several bulk quantities.\cite{Chandra02b}
In conventional charge- and spin-density waves (CDWs and SDWs respectively),
the quantity $A_{\bk }({\bf Q})$ is an isotropic
function of momentum. However in more general cases $A_{\bk }({\bf Q})$ will
develop a nodal structure which leads to anisotropy (Table
\ref{symmetries}) that is favoured by strong
Coulomb interaction, as we shall discuss in the next section.

\subsection{General Discussion of Anisotropic Charge Instabilities
                         in Fermi Liquids}\label{Landau}

At low temperatures, heavy electron materials form 
almost incompressible Landau Fermi liquids in which the residual
interactions between heavy quasiparticles are driven by strong,
low-lying antiferromagnetic spin fluctuations.  
This harshly renormalised
electronic environment is conducive to the development of
instabilities in which electrons or holes form bound-states
that contain nodes in their pair wavefunction.

Such arguments are well established in the context of anisotropic
Cooper pairing.\cite{Miyake86,Emery87} Here we extend these ideas, 
arguing that an almost incompressible
Fermi liquid is highly susceptible to the formation of 
anisotropic density waves, where the staggered 
electron-hole condensate contains a node in the pair wavefunction.
This issue
first arose in the context of orbital ordering in 
cuprate superconductors\cite{Nayak00}.   Here it has been emphasised
that strong Coulomb interactions suppress electron-hole bound-state
formation in CDWs, unless the bound-state contains a
node.\cite{Chakravarty01} 
Heavy electron fluids provide a unique opportunity to apply 
these arguments to three-dimensional systems.
Furthermore there is no controversy associated with the
Landau-Fermi liquid of their normal states, a situation in
distinct contrast to the situation in the cuprates.

In a heavy electron fluid, the density of states is severely
renormalised so that the ratio of the quasiparticle and the
bare band-structure density of states
\[
\frac{N^{*} (0)}{N (0)}\sim \frac{1}{Z}
\]
is typically at least a factor of ten. In these systems the 
magnetic susceptibility, given in Landau Fermi liquid theory 
by 
\[
\chi = \frac{N^{* (0)}}{1+ F_{o}^{a}}
\]
is weakly enhanced. By contrast, the charge susceptibility 
is severely depressed by strong coulomb interactions and
is essentially given by the unrenormalised band-structure
value
\[
\chi_{c} = \frac{N^{* (0)}}{1+ F_{o}^{s}}\sim N (0)
\]
which is why the fluid is characterised as ``almost incompressible''. 
It is this basic effect that 
rules out the formation of isotropic charge density
wave order and s-wave superconductivity.

Response functions that contain an anisotropic form
factor are unaffected by the strong Coulomb interactions. The key
point here is that 
the strong interaction effects are local and thus they do not affect
the higher Landau parameters, due to the nodes in the corresponding
spherical harmonics. 
For example
if we consider a ``chemical potential'' which couples anisotropically
to the Fermi surface in the l-th angular momentum channel, then the 
corresponding susceptibility is given by
\[
\chi ^{(l)}_{c}\sim \frac{N^{*} (0)}{1+F_{l}^{s}}\sim N^{*} (0)
\]
provided the higher Landau parameters are not much larger than unity. 
From this discussion, we see that 
large mass renormalisation and strong Coulomb repulsion
suppresses isotropic CDW formation
but that analogueous instabilities can form
in higher angular momentum channels.

\subsection{Anisotropic Pairings:  The Contenders}

We have just argued that the large Coulomb
repulsion between the heavy fermion quasiparticles (incompressibility)
in $URu_2Si_2$
discourages isotropic pairing in the CDW channel. 
This expectation is
confirmed by experiment, for charge density wave formation is 
expected to produce a 
lattice distortion, yet none is observed to develop
URu$_2$Si$_2$ below the 17K phase transition. 
Similarly neutron scattering is inconsistent with the
presence of an isotropic spin density wave in the
hidden ordered phase.\cite{Broholm87,Broholm91}
Thus, mainly due to the incompressibility of the heavy Fermi
liquid, we are left with three remaining anisotropic particle-hole
pairing states (see Table \ref{symmetries}).

The possibility of d-spin density waves as the hidden order in $URu_2Si_2$
has been raised by several 
authors.\cite{Ramirez92,Ikeda98}
In a Stoner analysis, d-SDWs require  {\sl ferromagnetic} exchange 
interactions of
neighbouring spins.
In particular, for antiferromagnetic interactions, 
a Stoner analysis reveals that the d-SDW
has a lower transition temperature than competing 
quadrupolar CDW (q-CDW) or spin-density waves.\cite{Kiselev99} 
Thus a d-SDW scenario favours ferromagnetic fluctuations in $URu_2Si_2$;
by contrast,
its transition at $T^* = 1.2 K$ to a d-wave superconductor indicates
the importance of antiferromagnetic fluctuations at $T > T^*$.

Before discussing the two remaining options presented within
the framework of Table \ref{symmetries}, we want to mention
two recent proposals for the hidden order parameter that
both lead to quasiparticle matrix elements similar to those
of a higher-order SDW.  
In the first one, the authors\cite{Mineev04} argue 
that consistency with experiment
can be maintained for an SDW that develops predominantly
in the p- or s- bands whose neutron form-factor at the Bragg peaks 
is significantly smaller than that of f-electrons.  
Here the key conceptual difficulty is that the matrix element of 
the order parameter in the f-bands would have to be small; yet the 
large entropy of condensation
observed at $T=T_0$ is almost certainly associated with
these same f-electrons. 
It has also been suggested\cite{Kiss04} 
the hidden order results from octupolar
crystal-field states. In the quasiparticle basis, such an order
parameter behaves like a spin-density wave with a small $g-$ factor.
At present, the viability of this approach awaits more detailed
predictions regarding the magnetic distributions within the
sample that then, like for the OAFM scenario, could be tested
by NMR and neutron measurements.

\figwidth=0.85\columnwidth

\fg{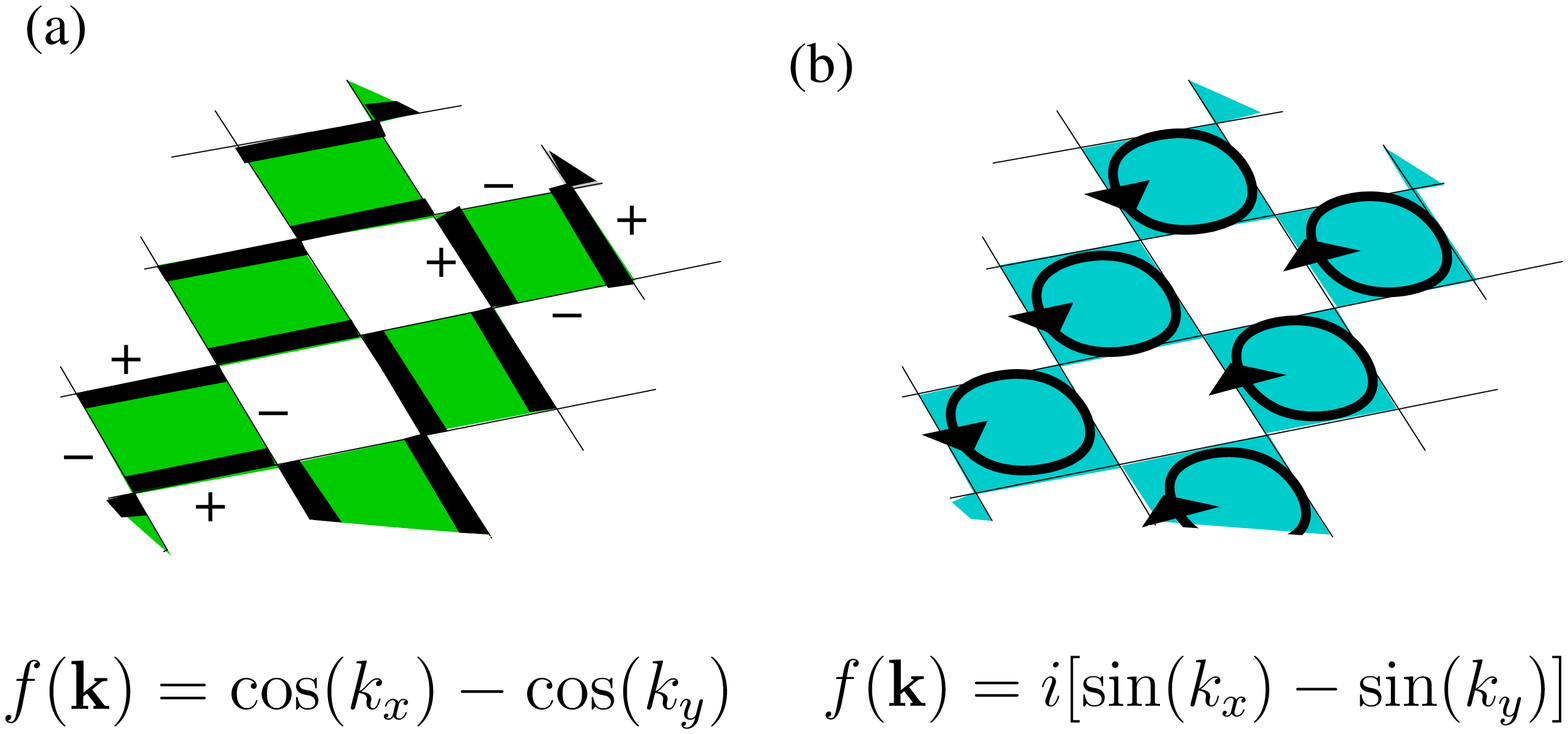}{scenarios}
{
(a) Incommensurate quadrupolar density wave (qCDW). In two
dimensions, the form factor $\cos (k_{x})-\cos (k_{y})$ leads to a
an incommensurate density wave with a quadrupolar charge distribution,
the CDW analogue of a d-wave superconductor.  (b) 
Incommensurate orbital antiferromagnet.  Here currents circulate
around square plaquettes defined by nearest-neighbour uranium ions.}

Returning to the table of possible pairing symmetries (Table {\ref{symmetries})
we therefore have two remaining options:  the quadrupolar charge density
wave\cite{Amitsuka02} (Fig. \ref{scenarios} (a))
 and the orbital 
antiferromagnet(Fig. \ref{scenarios} (b)).\cite{Chandra02a,Chandra02b}, where both scenarios
are consistent with our picture of $URu_2Si_2$ as an incompressible
Fermi liquid with strong antiferromagnetic fluctuations.  Each
order parameter has nodes, so that neither couples directly to the 
local charge density.  Furthermore both incommensurate density waves couples
weakly to uniform strain, and thus are both consistent with
the observed insensitivity\cite{Luthi93} of the elastic response at $T_0$.
Recent uniaxial stress measurements suggests that the hidden order is
sensitive to the presence of local tetragonal symmetry,\cite{Yokoyama02}
a feature that can be explained within both frameworks for completely
different reasons.  In the orbital antiferromagnet the currents
are equal in each basal direction\cite{Chandra02b}, whereas within
the quadrupolar scenario it is known that some of the crystal-field
states with tetragonal symmetry are quadrupolar.\cite{Santini00}  
Unfortunately the diamagnetic response cannot be used to discriminate
between these two scenario, as the contribution from orbital antiferromagnetism
is small compared to that associated with the gapping of the Fermi
surface ($\frac{\chi_{Pauli}}{\chi_{diam}} \sim 100$).

At present, the key factor distinguishing the orbital antiferromagnet
from the quadrupolar charge-density wave scenarios is the absence or
present of time-reversal breaking.  Because the local field distributions
and strengths measured by NMR have not yet been observed by other
methods, there is still uncertainty about these results.
We note that it has been argued\cite{Kiss04} that the observation
of a stress-induced moment\cite{Yokoyama02} implies that the
hidden-order breaks time-reversal symmetry; much as we would like
to believe this, we note that this result can be attributed to
a volume-fraction effect and thus is inconclusive.  
Both the quadrupolar charge density wave and the orbital antiferromagnet
have nodes in their respective gaps, which should in principle
be observable via photoemission and/or scanning tunnelling microscopy,
though issues associated with the nature of the surface of this
material remain to be resolved.  However the quadrupolar charge density
wave is not expected to lead to magnetic neutron scattering, and
therefore detailed elastic measurements are critical for resolving
the nature of the hidden order parameter.

\section{Toy model for Anisotropic Particle-Hole Pairing}\label{toymodel}

Next we explore a simple $t-J$ model for heavy electrons with
antiferromagnetic spin fluctuations, and explore different orderings.
We are motivated by experiment
in our choice of the model. 
URu$_{2}$Si$_{2}$ undergoes a phase transition to a 
d-wave superconducting state at $T_{0}=0.8 K$,  
and the pairing is understood to be mediated by antiferromagnetic
spin fluctuations. The same $t-J$ model also
encompasses orbital antiferromagnetism, quadrupolar CDW, and
isotropic SDW. 

   We consider a simplified model for the heavy Fermi liquid, described
by $H=H_{0}+H_{I}$, where
\[
H_{0}= \sum_{\bk}\epsilon_{\bk}c\dg_{\bk  \sigma}c_{\bk\sigma}
\]
describes the band of heavy electrons and 
\begin{equation}    
  H_{I}= \sum_{{\bf q}} J({\bf q})\, {\bf S}({\bf q}). {\bf S}({\bf -q}), 
\label{Hint}
\end{equation}
is the interaction between them. Here, 
${\bf S} (\bq ) = \frac{1}{2}c_{\bk +\bq\alpha }^{\dagger}{\pmb \sigma}_{\alpha
  \beta}c_{\bk \beta}$ is the Fourier transform of the local spin
operator. . 
In this simplified model, we consider  the 
indices $\sigma $ to represent  the pseudo-spin indices of the spin-orbit-coupled, heavy-electron
band. 
We recall that we are working in an itinerant basis where the local physics
(e.g. spin-orbit coupling) is absorbed into the composite quasiparticle states.
Using the completeness relation 
$\vec{\sigma}_{\alpha \beta }\cdot\vec{\sigma}_{\gamma \delta }+
\delta_{\alpha \beta}\delta_{\gamma \eta} = 2 \delta_{\alpha
\eta}\delta_{\gamma \beta }
$
we may rewrite this interaction as 
\[
H_{I}  = 
\frac{1}{2}\sum_{ij,\, \sigma \sigma '} J_{ij}\,
\left( c^{\dagger}_{i\sigma}
c_{i\sigma '}c^{\dagger}_{j\sigma '}c_{j\sigma}- \frac{1}{2}n_{i}n_{j}
\right). 
\]
Here we have rewritten the electron operators in a local basis, so
that 
$c_{j\sigma }= \frac{1}{\sqrt{N}} \sum_{\bk }c\dg
_{\bk\sigma}e^{i\bk\cdot {\bf x}_{j}}$ is the electron creation operator
at site $j$, $N$ is the number of sites in the lattice and $J_{ij}=
\frac{1}{N}\sum_{\bq }J (\bq )e^{i \bq \cdot (\bx_{i}-\bx_{j}}$ is the spin
interaction between sites $i$ and $j$. 
We shall  ignore the second term, which involves the heavily suppressed
fluctuations in quasiparticle occupation at each site. The first term
can be decoupled as 
\begin{eqnarray}
H_{I} 
 =  -\frac{1}{2N} \sum_{{\bf q, k, p},\sigma \sigma '}\!\!\!\!\!\!
J({\bf k}-{\bf p}) c^{\dagger}_{\sigma,\,{\bf k}_{+}}
c_{\sigma, \,{\bf k}_{-}}
c^{\dagger}_{\sigma ', \,{\bf p}_{-}}
c_{\sigma ',\,{\bf p}_{+}},
\label{Hint2}
\end{eqnarray}
where ${\bf k}_{\pm}={\bf k}\pm \frac{1}{2}{\bf q}$, etc.
The interaction potential $J({\bf q})$ can be expanded into partial waves,
\begin{equation}
V_{l} = 2\int_{0}^{1}dx\, xP_{l}(1-2x^{2})\,J(2p\,x),
\label{partial}
\end{equation}
where $x=\sin(\theta/2)$ and $p\approx p_{F}$. We require that the $l=0$
(isotropic) component be large and negative, reflecting strong on-site
quasiparticle repulsion. This has the effect of suppressing isotropic
particle-hole pairing. However this potential $V_{l}$, for $l>0$,
could be attractive,
which would then favour particle-hole pairing in higher angular
momentum channels. Such higher angular momentum components are present
due to anisotropy of the interaction $J({\bf q})$, which occurs 
at sufficiently large values of ${\bf q}$
where the underlying symmetry of the crystal becomes important. 

For the purposes of a toy model, 
we shall assume in URu$_{2}$Si$_{2}$, nearest neighbour antiferromagnetic spin
fluctuations (AFMSF) predominate, 
so that 
\begin{equation}
J({\bf q}) \approx 2\,J_{1}\gamma^{1}_{{\bf q}},
\label{interaction1}
\end{equation}
where the form factor $\gamma^{1}_{{\bf q}} =\cos(q_{x} a) + \cos(q_{y} a)$.   
With this approximation, 
the interaction in Eq.(\ref{Hint2}) is separable:
\begin{eqnarray}
H_{I}   =  
- \frac{J_{1}}{N}
 \sum_{{\bq \in {\frac{1}{2}\rm BZ}, k, p};\Gamma =1,4} (\gamma^{\Gamma}_{{\bf p}}\rho _{\bf p}({\bf q}))\dg \gamma^{\Gamma}_{{\bf k}}
\rho_{\bf  k}({\bf q}) 
 ,
\label{Hint3} 
\end{eqnarray}
where 
\begin{equation}
\rho_{\bf k}({\bf q})
=\sum_{\sigma} c^{\dagger}_{
{\bf
    k}+ \frac{1}{2}{\bf q}\sigma }c_{ {\bf k}-\frac{1}{2}{\bf
q}\sigma }
\label{phdensity}
\end{equation}
 are the particle-hole operators and 
\begin{eqnarray}\label{phpair}
\gamma^{1,2}_{\bf k}&=&\cos(k_{x}a) \pm \cos(k_{y}
    a)\cr\gamma^{3,4}_{\bf k} &=& i(\sin(k_{x}a) \pm \sin(k_{y}a)) 
\end{eqnarray}
are form factors that transform under the point-group symmetry of the
lattice. Since $\rho_{\bk } (\bq )=\rho\dg _{\bk } (-\bq )$, the
quantity inside the summation is symmetric under $\bq \rightarrow -\bq
$, and so, by doubling the prefactor and restricting
the sum over $\bq $ to one-half the Brillouin
zone, we  assure  that every term in the $\bq $ sum is independent. 

This  interaction is attractive and of equal magnitude
    in the four anisotropic channels. $\gamma^{1}_{{\bf k}}$,
    $\gamma^{2}_{{\bf k}}$, $\gamma^{3,4}_{{\bf k}}$  have {\it
    s}-like, {\it d}-like and {\it p}-like symmetry
    respectively.  
Notice that bond-variables
$\sum_{\sigma }\langle c\dg _{i\sigma }c_{j\sigma }\rangle$ are invariant under
time reversal, 
$\sum_{\sigma }\langle c\dg _{i\sigma }c_{j\sigma }\rangle =
\sum_{\sigma }\langle c\dg _{j\sigma }c_{i\sigma }\rangle^{*} $ and the 
imaginary pre-factors in 
$\gamma_{\bk }^{3,4}$ have been chosen so that the form-factors 
respect this symmetry, i.e 
$\gamma_{\bk }^{\Gamma}=
(\gamma_{-\bk }^{\Gamma})^{*}$.

By carrying out a ``Hubbard Stratonovich'' decoupling of $H_{I}$, we 
obtain
\begin{eqnarray}\label{Hmf1} 
H_{I}  & \rarrow  &
 \sum_{\bq \in \frac{1}{2}{\rm BZ},  \bk ;\Gamma =1,4} 
\left[\Delta^{\Gamma}_{\bq }
\gamma^{\Gamma}_{\bk} 
\rho_{\bk }({\bq})
+\bar \Delta^{\Gamma}_{\bq }
(\gamma^{\Gamma}_{\bk} )^{*}
\rho\dg _{\bk }({\bq})
 \right]\nonumber
\\
& +& \frac{N}{2J_{1}}
 \sum_{\bq\in {\rm\frac{1}{2}BZ};\ \Gamma =1,4}  
\bar \Delta^{\Gamma}_{\bq } \Delta^{\Gamma}_{\bq  }.
\end{eqnarray}  
Now the mean-field solution to this expression is determined by the
saddle-point condition 
\begin{equation}\label{gap}
\Delta^{\Gamma}_{\bq  } = - \frac{J_{1}}{N}\sum_{\bk } (\gamma^{\Gamma}_{\bk} )^{*}
\langle \rho_{\bk }(-{\bq})\rangle 
\end{equation}
In general, the density wave will condense
at a primary wavevector $\bq =\bQ$. For a
realistic model, $\bQ $ may well be incommensurate, in which case,
it will be accompanied by a family of corresponding $\bQ' $ that form
a ``star'' of q-vectors under the point group. There will in general
also be higher harmonics of $\bQ $. To illustrate the key ideas
however, we shall assume a simple model in which a single $\bQ $
dominates the density wave, i.e.
\[
\Delta_{\bq}^{\Gamma}=\Delta^{\Gamma}
\delta_{\bq ,\bQ}+
\bar \Delta^{\Gamma} \delta_{\bq ,-\bQ}
\]
For this discussion, we shall also assume that the Fermi surface is ``almost nested'', so
that the Fermi surface can be divided into two equal parts or reduced
Brillouin zones (RBZ): region I in which
$\vert \epsilon_{\bk -\frac{1}{2}\bQ}\vert \geq \vert \epsilon_{\bk +\frac{1}{2}\bQ}\vert 
$ and region II 
in which 
$\vert \epsilon_{\bk -\frac{1}{2}\bQ}\vert \leq
\vert \epsilon_{\bk +\frac{1}{2}\bQ}\vert 
$. 
In perfectly nested Fermi surfaces $\epsilon_{\bk }=
-\epsilon_{\bk +\bQ}$ are perfectly degenerate. For a square lattice and $\bQ= (\pi,\pi)$ the reduced Brillouin zone is
the diamond-shaped region bounded by $-\pi \leq k_{y} \leq \pi, -\pi +
|k_{y}|\leq k_{x}\leq \pi-|k_{y}|$.

The mean-field Hamiltonian is then
\begin{eqnarray}\label{l2}
H_{MFT} &=& 
\sum_{\bk \in RBZ} (c\dg _{\bk^{+}},c\dg_{\bk^{-}})
\left[
\begin{array}{cc}
\epsilon_{\bk^{+}}
& \Delta_{\bk }
\\
\bar \Delta_{\bk }& \epsilon_{\bk^{{-}}}
\end{array}
\right]
\left(
\begin{array}{c}
c_{\bk^{+}}\\
c_{\bk^{-}}
\end{array}
 \right)\cr
&+& N\sum_{\Gamma=1,4}\frac{
\bar \Delta^{\Gamma}\Delta^{\Gamma}
}{J_{1}}
\end{eqnarray}
where $\bk ^{\pm}=\bk\pm\bQ/2$ and $\Delta_{\bk }= \sum_{\Gamma}\Delta
^{\Gamma}\gamma^{\Gamma}_{\bk }$. 
Here,  
${\bf Q}$ is the wave vector for particle-hole pairing.
Diagonalising the electronic part of the total
Hamiltonian yields two bands,
\begin{eqnarray}
E_{{\bf k}}^{(\pm)} & = & \frac{1}{2}(\epsilon_{{\bf
    k}+\frac{1}{2}{\bf Q}}+\epsilon_{{\bf k}-\frac{1}{2}{\bf Q}})
    \nonumber \\
   &    &  \pm \sqrt{\frac{1}{4}(\epsilon_{{\bf k}
 +\frac{1}{2}{\bf Q}}-\epsilon_{{\bf k}-\frac{1}{2}{\bf
    Q}})^{2} + \vert \Delta_{{\bf k}}\vert ^{2}},
\label{bands}.
\end{eqnarray}

The mean field solution for pairing density $\langle \rho_{{\bf
    p}}({\bf Q}) \rangle $ in Eq.(\ref{gap}) is obtained by setting
the variation of the free energy $F$,
\begin{eqnarray}
F & = & - T\sum_{{\bf k}}^{RBZ} \ln [(1+\exp(-\beta E_{{\bf k}}^{(+)}))
(1+\exp(-\beta E_{{\bf k}}^{(-)}))]  \nonumber \\
&  & + N\sum_{\Gamma=1,4}\frac{
\bar \Delta^{\Gamma}\Delta^{\Gamma}
}{J_{1}}
\label{free_energy} 
\end{eqnarray}
with respect to $\bar \Delta^{\Gamma}$ to zero.
This yields the 
gap equation,  
\begin{eqnarray}\label{gapeqn}
\frac{\Delta^{\Gamma}}{J_{1}} = 
\frac{1}{N}
\sum_{{\bf k}, \Gamma}^{RBZ}
\Delta_{\bk } (\gamma_{{\bf k}}^{\Gamma})^{*}
\frac{f(E_{{\bf k}}^{(-)})-f(E_{{\bf k}}^{(+)})}{E_{{\bf
        k}}^{(+)}-E_{{\bf k}}^{(-)}}.
\end{eqnarray}
In the special case where condensation occurs in a single channel
$\Gamma=\Gamma_{0}$, this simplifies to
\begin{eqnarray}\label{gapz}
\frac{1}{J_{1}} = 
\frac{1}{N}
\sum_{{\bf k}}^{RBZ}
\vert \gamma_{{\bf k}}^{\Gamma_{0}}\vert^{2}
\frac{f(E_{{\bf k}}^{(-)})-f(E_{{\bf k}}^{(+)})}{E_{{\bf
        k}}^{(+)}-E_{{\bf k}}^{(-)}}.
\end{eqnarray}
At $T = T_0$, Equation (\ref{gapeqn}) is
essentially a Stoner criterion $J_1 \chi_{0\psi} (0) = 1$ where 
\begin{eqnarray}\label{susc}
\chi_{0\psi} ({\bf q}
) =\frac{1}{N}
\sum_{{\bf k}}^{RBZ}
\vert \gamma_{{\bf k}}^{\Gamma_{0}}\vert^{2}
\frac{f(\epsilon_{{\bf k}^{-}-\bq /2
})
-f(\epsilon_{{\bf k}^{+}+\bq /2})
}
{\epsilon_{{\bk}^{+}+\bq /2}-\epsilon_{{\bf k}^{-}-\bq /2}}, 
\qquad 
\end{eqnarray}
is the susceptibility associated with the hidden order parameter
$\psi$, 
measured at a wave vector ${\bf q}+{\bf{Q}}$. 

Without details of the band-structure we can not predict which of the four
order parameters will dominate. Some general comments are however in
order.  Although the pairing equation (\ref{gapeqn}) does not involve
any isotropic order
parameter, the extended-s wave order parameter $\gamma_{\bk}^{1}$
does have the same point-group symmetry as a pure s-wave and if it
condenses, will tend to induce charge modulation. In a real heavy
electron system, the effects of
Coulomb interaction will renormalise the effective coupling
constant for this channel, eliminating this order parameter from
consideration. Of the remaining cases, $\gamma_{\bk}^{2}$ corresponds
to a q-CDW order and $\gamma_{\bk}^{{3,4}}$ can be
associated with spontaneous orbital or line 
currents between the Uranium atoms, as we
shall now show. 

Let us consider the current 
\begin{equation}
j_{ij}= -\frac{iet}{\hbar}\sum_{\sigma }
(c^{\dagger}_{j\sigma }c_{i\sigma }-c^{\dagger}_{i\sigma }c_{j\sigma})
\label{curr}
\end{equation} 
from $i$ to $j$ along bond $i-j$. Orbital order corresponds to a
non-vanishing circulation of the current in a plaquette: 
\[I_{C}=\frac{1}{4a}\oint {\bf j}.d{\bf l}\neq 0\] 
and, therefore, is of the form (Fig. \ref{links1} (a) ).
\begin{eqnarray}
I_{C}({\bf X})  =  
\frac{1}{4}\left[ j_{12}+j_{23}+j_{34}+j_{41}\right],
\label{circul} 
\end{eqnarray}
where ${\bf X})$ 
is the position of the centre of the plaquette, and the indices
$(1-4)$ label the corners of the plaquette, taking the sense
of rotation to be anti-clockwise.

Now consider the evaluation of the bond variable $
\sum_{\sigma }\langle c\dg_{\sigma
} (\bx
+\ba /2)c_{\sigma } (\bx-\ba/2)
\rangle 
$. Taking the Fourier transform each electron field, we obtain
\begin{eqnarray}\label{l3}
&&\sum_{\sigma }\langle c\dg_{\sigma
} (\bx
+\ba /2)c_{\sigma } (\bx-\ba/2)
\rangle 
\nonumber
\\
&=& 
\frac{1}{N}\sum_{\bk ,\bk',\sigma}\langle c\dg
_{\bk\sigma}c_{\bk'\sigma}\rangle 
e^{i [\bk \cdot(\bx+ \ba /2) - \bk '\cdot(\bx-\ba/2)]}\cr
&=&\frac{1}{N}\sum_{\bk ,\bk'}\langle 
\rho_{{(\bk +\bk')}/{2}} (\bk -\bk')
\rangle 
e^{i [(\bk-\bk ')\cdot\bx
+ 
\overbrace {\scriptstyle
\frac{1}{2}(\bk +\bk')}^{\bp}
\cdot \ba ]}\cr
&=&
e^{i \bQ\cdot\bx
}\frac{1}{N }\sum_{\bp}
\langle 
\rho_{\bp } (\bQ)
\rangle 
e^{i 
 \bp\cdot \ba }+ (\rm \bQ \leftrightarrow - \bQ )
\end{eqnarray}
where we assumed $\langle \rho_{\bk} (\bq )\rangle = 
\delta_{\bk ,\bQ}\langle \rho_{\bk} (\bQ )\rangle + (\bQ
\leftrightarrow -\bQ ) $. The current along a given bond is therefore
\begin{eqnarray}\label{l4}
&&j (\bx+\ba /2 , \bx -\ba/2 )\cr
&=& e^{i \bQ\cdot\bx
}\frac{et}{N\hbar }\sum_{\bk}
\langle 
\rho_{\bk } (\bQ)
\rangle 
2 \sin ( 
 \bk\cdot \ba )+ (\rm H.c)
\end{eqnarray}
Averaging the currents anti-clockwise around a plaquette centred at $\bX$, we 
arrive at 
\begin{equation}
I_{C}({\bf X})= I_{C}\exp[i{\bQ}.{\bf X}]+ {\rm H.c}
\label{curr2z}
\end{equation}
where
\begin{eqnarray}\label{l5}
I_{C} ({\bf X}) &=&
i\frac{et}{N\hbar }
\sum_{{\bf k}}\langle \rho_{\bk} (\bQ )\rangle 
[
s_{x}
\sin ( Q_{y}a/2)
-s_{y}\sin ( Q_{x}a/2)\cr
&=&
\frac{et}{N\hbar }
\sum_{{\bf k}}\langle \rho_{\bk} (\bQ )\rangle 
[\alpha_{+}\gamma^{4}_{\bk }+
\alpha_{-}\gamma^{3}_{\bk }
],
\label{curr2}
\end{eqnarray}
where we have used the notation $s_{x,y}\equiv \sin(k_{x,y}a)$ and
\[
\alpha_{\pm}= \frac{1}{2}\left[
\sin ( Q_{y}a/2)
\pm\sin ( Q_{x}a/2)
\right]
\]
(notice the ordering of the y and x terms).
The form factor for orbital current order is thus a weighted mixture
of $\gamma_{{\bf k}}^{4}$ and $\gamma_{{\bf k}}^{3}$. 
Using Eq.(\ref{gap}) to simplify 
Eq. (\ref{curr2}), we obtain a relation between the orbital current and
gap,
\begin{equation}
I = \frac{e\Delta_{C}}{\hbar 
}. \frac{t}{J_{1}}.
\label{gapcurrentrel}
\end{equation}
where 
\[
\Delta_{C} = 
\alpha_{+}
\Delta^{4}+\alpha_{-}
\Delta^{3}.
\]
In actual fact, the relative weight of the two channels in the
orbital antiferromagnet is not an adjustable parameter. If we
calculate the divergence of the current at a given node in the
lattice, we find that
\begin{eqnarray}\label{l6}
&&\nabla \cdot {\bf{ j}} (\br) \cr
&=& j (\br + a \hat \bx ,\br )
-j (\br - a \hat \bx ,\br ) +j (\br + a \hat \by ,\br )-
j (\br + a \hat \by ,\br )\cr
&=& \frac{4 e t}{\hbar J_{1}}\left(
\alpha_{+}
\Delta^{3}-\alpha_{-}
\Delta^{4}
 \right)=0
\end{eqnarray}
so the choice of $\bQ $ vector determines the mix of $\gamma^{3}$ and
$\gamma^{4}$ symmetry in the orbital antiferromagnet. 

In an itinerant model the condition for instability into the
hidden order phase will be given by the Stoner criterion
already discussed ($J_1 \chi_{0\psi} = 1$) where 
$\chi_{0\psi} \sim \frac{1}{t}$
so that typically at the transition $I = \frac{\beta e\Delta}{\hbar}$
where $\beta \sim O(1)$ is a constant; this is the relation we used
for the current in our earlier phenomenological treatment.  
The form factor $\gamma_{{\bf k}}^{2}=\cos(k_{x})-\cos(k_{y})$
corresponds to a quadrupolar charge density wave (q-CDW). The
particular details of the conduction electron spectrum 
$\epsilon_{{\bf k}}$ determine which order parameter has a higher
critical temperature.
For instance, if $(Q_{x},Q_{y})=(\pi,\pi)$, and $\epsilon_{{\bf
   k}}=-t(\cos(k_{x})+\cos(k_{y}))$, which corresponds to
a nested Fermi surface, then from 
Equation (\ref{gapz}) the relation for $T_{0}$ is 
\begin{eqnarray}
1 & = &
\frac{J_{1}}{8\pi^{2}t}\int_{-\pi}^{\pi}dk_{y}
\int_{-\pi+|k_{y}|}^{\pi-|k_{y}|}dk_{x}| \gamma_{{\bf k}}^{i}|^{2}
\frac{\mbox{tanh}\left(\frac{E_{{\bf
      k}}}{2T_{0}^{i}}\right)}{E_{{\bf k}}}.
\label{eqforTo}
\end{eqnarray}
Here we used 
$E_{{\bf k}}(\Delta=0)\equiv\epsilon_{{\bf k} + {\bf Q}/2}=
 t(\sin(k_{x})+\sin(k_{y}))$. 
In this particular limit, we can explicitly 
verify that orbital antiferromagnetism has a higher $T_{0}$ than
 q-CDW. 
In the real material, the spectrum may
 differ greatly from this simple form, which may result in a preference of
 q-CDW over OAFM.

Our discussion in this section is based on a weak-coupling treatment of
orbital antiferromagnetism, which is technically only valid in the
vicinity of a nesting instability.  Real heavy
electron systems involve interactions of a size comparable with the
band-width, in which the vicinity to nesting will no longer be a
requirement. 
Practical modelling of these situations will require
alternate strong-coupling methods, such as methods based on a Kondo
lattice model.  It is however interesting to note that 
that both symmetry and microscopic
toy treatments appear to point to quadrupolar charge density wave and orbital antiferromagnetism as the leading contenders for hidden order in $URu_2Si_2$.
 
\section{Fluctuations and Nesting}\label{sharpness}

The sharpness of the phase transition in URu$_2$Si$_2$ indicates
that fluctuations do not make a significant contribution to
thermodynamic properties.  From the observed specific heat anomaly,
the region of fluctuations is certainly smaller than $\Delta T \sim 0.1K$, 
so that $t_{g}= \Delta T/T_{0} <  \frac{1}{200}$.  This is an unusual
situation in the general context of local moment magnetism, where
broad fluctuation regions are generally seen in the specific heat
anomaly.
This result is sometimes taken to indicate that the hidden order involves 
a nested Fermi surface.\cite{Ikeda98,Sikkema96}  
However band structure calculations have not revealed
any signs of a nested Fermi surface,\cite{Norman88} and it is difficult
to see how such a condition might occur naturally in the
complex band-structure of an f-electron system. 
Sharp mean-field transitions are generally taken as an indication of a large
coherence length scale associated with fluctuations. In insulating
systems (e.g. ferroelectrics) this arises from the long-range
nature of the interaction. In superconductors and in nested charge
density wave systems, the long coherence length $\xi_{0}=v_{F}/\Delta
$ is a consequence  of the  non-local order parameter response of the
itinerant electron fluid. 

So can the sharpness of the specific heat
transition can be used in $URu_{2}Si_{2}$
be used to infer the presence of nesting $URu_{2}Si_{2}$?  
In fact, as we shall now see, a careful
examination of the Ginzburg criterion for this system shows that while
we may confirm that the ordering is itinerant in nature, the small
size of the heavy electron Fermi energy means that we do not need to invoke
nesting to understand that sharpness of the transition.  

The
Ginzburg criterion for a phase transition is given by 
\begin{equation}
   t_{G}=\frac{1}{[(\xi_{0}/a)^{d}(\delta S/k_{B})]^{2/(4-d)}},
\label{ginzburg1}
\end{equation}
or in  three dimensions,
\begin{equation}
   t_{G}=\frac{1}{(\xi_{0}/a)^{6}(\delta S/k_{B})^{2}},\qquad \qquad (d=3)
\label{ginzburg2}
\end{equation}
Here $a$ is the lattice spacing,   $\delta S$ is the entropy associated
with the phase transition and $\xi_{0}$ is the coherence length 
of the order parameter. 
Microscopically, $\xi_{0}$ is determined from the Gaussian
fluctuation term in the order-parameter expansion of the free energy,
\begin{equation}\label{freeform}
\Delta F \sim \frac{1}{2}\int_{\bq } \alpha \vert \Psi_{\bq}\vert^{2} (t +
q^{2}\xi_{0}^{2}) 
\end{equation}
where $t= \frac{T}{T_{0}}-1$ and $\alpha $ is a normalisation
constant. The Gaussian coefficient  in the integral is directly
related to the static susceptibility of the order parameter
\begin{equation}\label{}
\chi_{\psi }^{-1} (q) =\alpha  (t+ q^{2 }\chi_{0}^{2})
\end{equation}
The relationship between the coherence length $\chi_{0}$ and
microscopic quantities depends markedly on the underlying physics.
In  insulators, $\xi_{0}$ tends to be determined by the range of interaction
of the order parameter, but in itinerant systems, it is determined by
the non-local order-parameter polarisation that develops in the
electron fluid. 

For example, in a local-moment antiferromagnet, with
interaction $H= \frac{1}{2}\sum_{q}S_{q}S_{-q}$, 
\begin{equation}\label{}
\chi_{q}^{-1} = \mu_{B}^{-2} (T + J_{q})
\end{equation}
When we expand around the unstable $q$ vector, $q=Q_{0}$
\begin{equation}\label{}
J (q) = \theta_{C} (1- \kappa^{-2}(\vec{q}-\vec{Q}_{0})^{2})
\end{equation}
where $\theta_{C}=-T_{0}$ is the Curie constant and $\kappa^{-1}$ the
effective range of the interaction. With this form, we see that for
insulating systems, the coherence length 
$\xi_{o}=\kappa^{-1}$ becomes the range of the interaction. For
short-range interactions, this reason, the breadth of fluctuation
region  is generally large. In insulating systems, narrow fluctuation
regimes are therefore associated with long-range interactions. 

By contrast, in itinerant electron systems the 
order-parameter susceptibility generally takes the form
\begin{equation}\label{}
\chi_{\psi}^{-1} (q) = -g + [\chi_{0\psi} (q)]^{-1}
\end{equation}
where $g$ is the strength of short-range interaction between electrons in the
channel corresponding to the order parameter and $\chi_{0\psi}$
takes the form given in (\ref{susc}).
It is the momentum dependence of $\chi_{0} (q)$ that determines the
Ginzburg criterion in itinerant systems.
To understand the role of nesting, let us consider a Fermi surface in which the departure from dispersion is
measured by an energy scale $\mu$, (e.g $\epsilon_{\bk }=  -2 t (\cos
k_{x}+\cos k_{y})-\mu$) then the dispersion satisfies
\begin{equation}\label{}
-\epsilon_{\bk-{\bf Q} }=  \epsilon_{\bk}+ 2\mu
\end{equation}
so that the bare susceptibility (\ref{susc})
is given by
\begin{eqnarray}\label{suscb}
\chi_{0\psi} ({\bf q}
) =\sum_{{\bf k}}^{RBZ}
\vert \gamma_{{\bf k}-{\bf Q}/2}^{\Gamma}\vert^{2}
\frac{f(\epsilon_{{\bf k}^{-}})
-f(\epsilon_{{\bf k}^{+}})
}
{\epsilon_{{\bk}+\bq /2}-\epsilon_{{\bf k}-\bq /2}+2\mu}, 
\end{eqnarray}
For $\mu=0$ and $q=0$, this integral is logarithmically divergent at
$T=0$, and given by $\chi_{0\psi } (q=0)\sim \rho \bar {\vert
\psi^{\Gamma}\vert^{2}}\ln \left(\frac{D}{T} \right)$ at finite
temperatures. 
Finite $\bq $ modifies the Fermi functions in
(\ref{suscb}), so that
\begin{eqnarray}\label{approxy}
 [\chi_{\psi}^{0} (q)]&\sim & \left(1 - \frac{(v_{F}q)^{2}}{4}
\frac{\partial^{2}}{\partial T^{2}}\right)\chi_{0\psi } (T)
\cr
&\sim& \rho \left(\ln \left(\frac{D}{T} \right) - \frac{(v_{F}q)^{2}}{T^{2}}\right)
\end{eqnarray}
so that 
\begin{equation}\label{}
- g + \chi_{0\psi }^{-1}= \frac{g}{\ln (D/T_{0})}\left[ t + \left(\frac{v_{F}q}{T_{0}} \right)^{2} \right]
\end{equation}
and the
free energy expansion takes the form
\begin{equation}\label{l7}
\Delta F \sim \int_{\bq } \vert \Psi_{\bq}\vert^{2}
\left(\left(\frac{\delta T}{T_{0}} \right) +\left(\frac{v_{F}q}{T_{0}} \right)^{2}
\right)^{2})
\end{equation}
so by comparing with (\ref{freeform}), we see that 
for a nested system, the coherence length takes the ``BCS'' form
\begin{equation}\label{}
\xi_{0}\sim \frac{v_{F}}{T_{0}}.
\end{equation}
When $\vert \mu \vert >>
T_{0}$, then we must replace $T_{0}\rightarrow \vert  \mu \vert $ in
the Landau-Ginzburg expansion, i.e.
\begin{eqnarray}\label{l8}
\Delta F &\sim & 
\int_{\bq } \vert \Psi_{\bq}\vert^{2}
\left(\left(\frac{\delta T}{\vert \mu\vert } \right)
+\left(\frac{v_{F}q}{\mu} \right)^{2}
\right)^{2} \cr
&=& \left(\frac{T_{0}}{\vert \mu\vert } \right)\int_{\bq } \vert \Psi_{\bq}\vert^{2}
\left(\left(\frac{\delta T}{T_{0} } \right)
+\left(\frac{v_F q}{\vert \mu \vert T_{0}}\right)^{2}\right)^{2}
\end{eqnarray}
from which we see that the coherence length is given by
\begin{equation}\label{}
\xi_{0}\sim \frac{v_{F}}{\sqrt{T_{0}\vert \mu\vert }}\sim
\sqrt{\xi_{nested}a}
\end{equation}
where we have replaced $\frac{v_{F}}{\vert \mu \vert }\sim a$, so
loosely speaking, the absence of nesting replaces the coherence length by
the geometric mean of the BCS coherence length $v_{F}/T_0$ and the
lattice spacing. 

Let us now return to our case, $URu_{2}Si_{2}$. Here, using the three
dimensional form of the Ginzburg criterion, and taking 
$URu_{2}Si_{2}$, $\delta S\sim 0.1 k_{B}$, 
so that 
\begin{equation}
   t_{G}\sim\frac{100}{2(\xi_{0}/a)^{6}}.
\label{ginzburg3}
\end{equation}
Suppose the fluctuation region is less than $0.1K$,
i.e. $t_{G} < (0.1K/20K)\sim 1/200$, then a lower bound for the
coherence length is 
\[
\xi_{0}/a \sim (t_{G}/100)^{-\frac{1}{6}}= (2\times 10^{4})^{1/6} \sim 5
\]
Clearly, the presence of the sixth power in the Ginzburg
criterion, means that only  modest coherence  length is  required to
account for experiments. 
Were the hidden order strictly associated with the local moments, 
then we would expect  $\xi_{0}/a\sim 1$, and clearly, the absence of
fluctuations is sufficient to rule this case out. The high pressure
magnetic phase transition does in fact show
clear signs of Ising fluctuations, and in this region, it would appear
that the ordering transition is indeed local in nature.
However, we can account for the coherence
length of the hidden order transition by appealing to itineracy,
without nesting. By assuming that
$v_{F}/\Delta\sim\frac{\epsilon_{F}}{\Delta }a \approx 25a$.  Taking
$\epsilon_{F}\sim 10^{3}K$, consistent with the heavy mass
$m^{*}/m_{e}\sim 60$ 
and $\Delta \sim 10^{2}K$, we are clearly in the
right range. 
From these arguments,
we see
that a correlation length of order $5$ lattice spacings is fully
consistent with a system that is  un-nested but itinerant.   We conclude
that the sharpness of the hidden order phase transition in
$URu_{2}Si_{2}$ only implies itineracy.
Indeed, there
are a number of heavy electron systems with sharp thermodynamic transitions
and commensurate magnetic order, indicative of un-nested Fermi
surfaces, such as 
$U_{2}Zn_{17}$\cite{Ott84} and $UPd_{2}Al_{3}$.\cite{Geibel91} 
In each of these cases, it is most
likely the itineracy alone that is responsible for the narrow fluctuation regime.

\section{Discussion}\label{discussion}
The observed Fermi liquid behaviour for $T > T_0$, the sharp nature of the transition
and the large entropy loss point to the hidden order as a general density-wave
with itinerant excitations formed from the local spin and orbital degrees 
of freedom of the uranium ions and f-electrons.  Motivated by nuclear magnetic
resonance measurements, we have expanded on our proposal (with J.A. Mydosh)
of the hidden order as incommensurate orbital antiferromagnetism and have provided technical
details for our predictions for elastic neutron scattering.  Next we have
turned to a microscopic description of the hidden order.  After discussing
symmetries and allowed particle-hole pairings in general terms, we studied 
the developing of these ordering in the setting of a toy single-band
$t-J$ model within a weak-coupling approach. 
Within this framework,
selection between q-CDW and OAFM ordering is not possible, though the situation 
may be different in the (experimentally relevant) strong-coupling regime.
As discussed in
Sec.\ref{Landau}, density wave
instabilities such as q-CDW and OAFM can account for the large entropy
loss observed at the transition ($\delta S\approx k_{B}^{2}T_{0}N^{*}(0)$) 
if the density of states at the Fermi surface, $N^{*}(0)$, is large (as is the case in a
heavy Fermi-liquid), and there is a substantial gapping of
the Fermi surface.

The weak-coupling model we considered requires the nesting
of a significant part of the Fermi surface. 
This requirement can be relaxed if the coupling is strong. 
Indeed it seems that a strong coupling description
might be more appropriate for URu$_2$Si$_2$, since the transition
temperature $T_{0}$ is an order of magnitude smaller than the
gap $\Delta$ unlike a weak coupling description where $T_{0}$
is more comparable with $\Delta$.  
Unfortunately here it is difficult to perform controlled calculations in this
regime,
and thus experiment is crucial for discerning between these two competing
scenarios of quadrupolar charge density wave order and orbital antiferromagnetism.
In
Sec. II we studied the consequences of OAFM for the
neutron scattering structure-factor $\mathbf{S}(\mathbf{q})$ and NMR at the
Si and Ru sites. No particular microscopic model was assumed here, so
the analysis is applicable for any coupling. NMR observations were
used with our OAFM model to predict an incommensurate wavevector for
orbital ordering which may be verified by neutron scattering
measurements. To date, these predictions remain untested, as current
experimental resolution is insufficient to observe the anticipated signal
level from an OAFM.\cite{Bull02,Wiebe04}  
Here we identify a region of momentum space where 
elastic neutron scattering probes will clearly be able to distinguish
between a spin density wave and an OAFM with current signal-noise levels.
This prediction for orbital anferromagnetism remains a challenge for
future experiments.   

Our proposal of orbital antiferromagnetism is strongly motivated by
the inhomogeneous line-broadening
observed in ambient pressure NMR,\cite{Bernal01} and there are questions 
associated with this experiment that concern us greatly. In particular,
the local fields measured via NMR in epoxied powdered samples
are an order of magnitude larger than those probed by muon spin
resonance or nuclear magnetic resonance in single-crystal ones.
One interesting possibility is ``motional narrowing''.
The proposed orbital antiferromagnetic order is incommensurate and 
quite similar in its current patterns to a flux lattice of core-less vortices,
where the absence of vortex cores weakens the pinning effect of
disorder.  In single domain crystals
an incommensurate orbital antiferromagnet should then be weakly
pinned, giving rise to large
thermal
motion.\cite{Blatter94}  
The probed local-fields
will then be ``motionally narrowed'', i.e their time-average
will be significantly reduced in magnitude relative to their its static counterpart.  
One of the predictions of this scenario, is that the 
the muon or NMR linewidth will increase systematically with disorder - 
an effect that might be tested using radiation damage to
systematically tune the disorder in a single sample.

Since we started working on this project, there have been a number of
new experiments which may place further constraints on the nature
of the hidden order in $URu_2Si_2$.  In particular, recent magnetotransport
measurements\cite{Bahnia05} indicate an unusually large Nernst signal
in $URu_2Si_2$ that develops at $T= T_0$. This kind of behaviour has
also been seen in the pseudogap phase of 
underdoped cuprate superconductors.  In the case of the cuprate
superconductors, this is most likely an effect of the Magnus force on
the pre-formed pairs in the pseudogap.  However, here, the absence of
any superconductivity makes it far more likely that the giant Nernst
effect seen here is a property of the 
quasiparticles in the presence of 
the hidden order parameter.   These new results clearly place an important
constraint on the microscopic nature of the order parameter. 

Recent high magnetic field studies\cite{Kim03} have raised additional
questions about the hidden order in $URu_2Si_2$. Application of high
magnetic fields confirms that the hidden order persists to significantly
higher values than does the remnant antiferromagnetism, affirming
the two-phase scenario.\cite{Chandra02a} Moreover the application
of still higher fields leads to a profusion of new hidden order phases
that may well  cloak a field-induced  quantum critical point. 
At the current time, it is not yet clear whether the proposed quantum
critical point is a consequence of the loss of hidden order, or
whether it might arise from the close vicinity to a quantum critical
end point (as is the case\cite{Millis02} with $SrRu_{2}O_{4}$.)

The hidden order mystery in Uranium Ruthenium-2 Silicon-2 can be
regarded as part of a much broader set of longstanding problems that our
community faces in the context of highly correlated materials.
Coexisting forms of hidden order, novel metallic states manifested by unusual
resistance and magnetotransport properties, field-induced quantum phase transitions-
each of these features,  present  in $URu_{2}Si_{2}$, are 
manifest themselves in a wide range of other strongly correlated
materials, such as the cuprate superconductors, strontium ruthenate, 
magneto-resistance materials, and many other heavy electron systems.
$URu_{2}Si_{2}$ offers an alternative perspective on these problems,
and optimistically, its ultimate solution will provide part of the key
to understanding these broader questions.

\appendix

\section{Spatial distribution of vector potential due to OAFM}
\begin{widetext}
Here we calculate the vector potential  ${\bf A}({\bf x})$ due to orbital
order by summing up contributions from currents in all
links. Consider first the contribution ${\bf A}^{12}$  
defined in Eq.(\ref{A}) for links along $\langle 12 \rangle$.
Performing the integral over $w$ gives
\begin{eqnarray}
{\bf A}^{12}({\bf x}) & = & {\hat {\bf x}}\frac{I_{0}}{c}\sum_{j}
e^{-i{\bf Q}.{\bf X}_{j}} \left[\sinh^{-1}\left(\frac{a/2 +
  X_{j} -x}{\sqrt{(y-Y_{j}+a/2)^{2}+(z-Z_{j})^{2}}}\right)\right.-
\nonumber \\ 
   &  & -\left.\sinh^{-1}\left(\frac{-a/2+X_{j}-x}{\sqrt{(y-Y_{j}+a/2)^{2}+(z-Z_{j})^{2}}}\right)\right].
\label{A12}
\end{eqnarray}
where we have used the notation ${\bf X}_{j}= (X_{j},Y_{j},Z_{j})$ to
denote the co-ordinates of the centre of the plaquette. 
For the link $\langle 43\rangle$ shown in Fig.\ref{links1} we have
\begin{eqnarray}
{\bf A}^{43}({\bf x}) & = & -{\hat {\bf x}}\frac{I_{0}}{c}\sum_{j}
e^{-i{\bf Q}.{\bf X}_{j}} \left[\sinh^{-1}\left(\frac{a/2 +
  X_{j} -x}{\sqrt{(y-Y_{j}-a/2)^{2}+(z-Z_{j})^{2}}}\right)\right.-
\nonumber \\ 
   &  & -\left.\sinh^{-1}\left(\frac{-a/2+X_{j}-x}{\sqrt{(y-Y_{j}-a/2)^{2}+(z-Z_{j})^{2}}}\right)\right].
\label{A43}
\end{eqnarray}
The $x$ component of the vector potential is then $A_{x}({\bf x})=
A^{12}({\bf x})+ A^{43}({\bf x})$. Similarly the vector potential in
the links $\langle 14 \rangle$ and $\langle 23 \rangle$,
\begin{eqnarray}
{\bf A}^{14}({\bf x}) & = & -{\hat {\bf y}}\frac{I_{0}}{c}\sum_{j}
e^{-i{\bf Q}.{\bf X}_{j}} \left[\sinh^{-1}\left(\frac{a/2 +
  Y_{j} -y}{\sqrt{
(x-X_{j}+a/2)^{2}+(z-Z_{j})^{2}}}\right)
\right.
-
\cr
   &  & -\left.\sinh^{-1}\left(
\frac{-a/2+Y_{j}-y}{\sqrt{(x-X_j}+a/2)^{2}+(z-Z_{z})^{2}}\right)\right],
\label{A14} \\
{\bf A}^{23}({\bf x}) & = & {\hat {\bf y}}\frac{I_{0}}{c}\sum_{j}
e^{-i{\bf Q}.{\bf X}_{j}} \left[\sinh^{-1}\left(\frac{a/2 +
  Y_{j} -y}{\sqrt{(x-X_{j}-a/2)^{2}+(z-Z_{j})^{2}}}\right)\right.-
\nonumber \\ 
   &  & -\left.\sinh^{-1}\left(\frac{-a/2+X_{jy}-y}{\sqrt{(x-X_{j}-a/2)^{2}+(z-Z_{j})^{2}}}\right)\right],
\label{A23}
\end{eqnarray}
yield the $y$ component of the vector potential $A_{y}({\bf x})=
A^{14}({\bf x})+ A^{23}({\bf x})$. The magnetic field follows
straightforwardly from ${\bf B}= {\bf \nabla}\times {\bf A}$.
\end{widetext}

\acknowledgments
We are very grateful to J.A. Mydosh for innumerable discussions on $URu_2Si_2$
which have shaped our view on this subject.
We have also benefitted greatly from interactions with W.J.L. Buyers 
particularly
related to the experimental neutron scattering situation and the Ginzburg
criterion in itinerant systems.  Discussions with 
K. McEwen, B. Maple, B. Marston, P. Ong and I. Usshikin are also acknowledged. 
VT is supported by a Junior Research Fellowship from Trinity College,
Cambridge. P. Chandra and P. Coleman are supported by the grants 
NSF DMR 0210575 and NSF DMR 0312495 respectively. P. Coleman acknowledges
the hospitality of the Kavli Institute for Theoretical Physics, where
part of this work was performed.

\end{document}